\documentclass[preprint,twocolumn]{aastex631}
\usepackage{subfigure}
\usepackage{color}
\usepackage{xcolor}
\usepackage{amsmath}
\usepackage{mathrsfs}
\usepackage{multirow}
\usepackage{soul}
\usepackage{amssymb}
\usepackage{hyperref}
\definecolor{ultramarine}{rgb}{0.8, 0.1, 0.4}

\def\be{\begin{equation}}
\def\ee{\end{equation}}
\def\bd{\begin{displaymath}}
\def\ed{\end{displaymath}}
\def\ba{\begin{aligned}}
\def\ea{\end{aligned}}

\def\nms{\mathsurround=0pt}

\def\oversim#1#2{\lower 4pt\vbox{\baselineskip 0pt \lineskip 1pt
    \ialign{$\nms#1\hfil##\hfil$\crcr#2\crcr\sim\crcr}}}
\def\ga{\mathrel{\mathpalette\oversim>}}

\def\bh{M_{\bullet}}

\def\msun{M_{\odot}}

\def\kms{{\rm km\,s^{-1}}}

\def\pl{\partial}

\def\GNC{\texttt{GNC}}
\begin{document}

\title{Monte-Carlo Stellar Dynamics near Massive Black Holes\\ 
Two-dimensional Fokker-Planck solutions of multiple mass components.}
\author{Fupeng Zhang}
\correspondingauthor{FUPENG ZHANG}
\affiliation{School of Physics and Materials Science, Guangzhou
University, Guangzhou 510006, China}[0]
\email{zhangfupeng@gzhu.edu.cn}
\affiliation{Key Laboratory for Astronomical Observation and Technology of Guangzhou, 510006 Guangzhou, China}
\affiliation{Astronomy Science and Technology Research Laboratory of Department of Education of Guangdong Province, Guangzhou 510006, China}
\author{Pau Amaro Seoane}
\affiliation{Institute of Multidisciplinary Mathematics, Universitat Politècnica de València, Spain}
\affiliation{
Max-Planck-Institute for Extraterrestrial Physics, Garching, Germany}
\affiliation{
The Higgs Centre for Theoretical Physics, University of Edinburgh, UK
}
\affiliation{Kavli Institute for Astronomy and Astrophysics, Beijing, China}

\begin{abstract}
%In this work, we have developed a new Monte-Carlo code (\GNC) that can obtain the dynamical relaxation 
%of a cluster consisting of multiple mass components near the massive black hole in center of galaxies.
%The method is based on two-dimensional (in space of energy and angular momentum) Fokker-Planck equations, and for the first time, 
%can evolve multiple mass components for particles of different types, including stars and compact objects. 
%The code is very flexible to include other complex dynamics, e.g., the resonant relaxations, 
%the gravitational wave orbital decay, and etc, in the future. The code has adopt 
%a weighting method that can increase significantly the statistics of the results of rare particles. 
%In this first paper we describe the basic version of our method, which consider only the two-body relaxations and the 
%effects of loss cone. We compare the results from \GNC{} with those from previous %studies and 
 %find general consistencies in the relaxation processes, energy and angular momentum distributions, density profiles and loss cone consumption rates.
%We find that a tangential anisotropy is always developed inside the cluster, 
%although at outer radius the cluster can remain nearly isotropic. 
%In the future \GNC{} can be applied to investigate a broad number of interesting phenomena in the galactic nuclei 
%and can obtain results with great details.
In this study we present a novel Monte-Carlo code, referred to as \GNC, which enables the investigation of dynamical relaxation in clusters comprising multiple mass components in the vicinity of supermassive black holes at the centers of galaxies. Our method is based on two-dimensional Fokker-Planck equations in the energy and angular momentum space, and allows the evolution of multiple mass components, including stars and compact objects. The code demonstrates remarkable flexibility to incorporate additional complex dynamics, such as resonant relaxations and gravitational wave orbital decay. By employing a weighting method, we effectively enhance the statistical accuracy of rare particle results. In this initial publication, we present the fundamental version of our method, focusing on two-body relaxations and loss cone effects. Through comparisons with previous studies, we establish consistent outcomes in terms of relaxation processes, energy and angular momentum distributions, density profiles, and loss cone consumption rates. We consistently observe the development of tangential anisotropy within the cluster, while the outer regions tend to retain near-isotropic characteristics. Moving forward, \GNC{} holds great promise for exploring a wide range of intriguing phenomena within galactic nuclei, in particular relativistic stellar dynamics, providing detailed and insightful outcomes.
\end{abstract}

\keywords{Black-hole physics -- gravitation -- Galaxy: center -- Galaxy: nucleus -- relativistic processes -- stars:
kinematics and dynamics }

\section{Introduction}

It is widely believed that massive black holes (MBHs) reside in the center of most galaxies, embedded by dense clusters consisting of stars and compact objects (the so-called nucleus star clusters). The deep potential of the MBH significantly changes the stellar dynamics, and the strong interplay between the stars and the MBH leads to a number of violent phenomena in galactic nuclei.

Many of these phenomena are tightly connected to the evolution of MBHs and the star formation history of galactic nuclei, e.g., the tidal disruption of stars \citep[e.g.,][]{1988Natur.333..523R}, ejection of hyper-velocity stars \citep[e.g.,][]{1988Natur.331..687H}, and stellar collisions \citep[e.g.,][]{1983ApJ...268..565D,1990ApJ...356..483Q}.

Others are important sources of gravitational wave events, such as extreme-mass ratio inspirals (EMRIs) \citep[e.g.,][]{1997MNRAS.284..318S,2003ApJ...583L..21F,2005ApJ...629..362H,2013MNRAS.429.3155A,2014MNRAS.437.1259B, 2018LRR....21....4A,2019PhRvD..99l3025A,2022hgwa.bookE..17A}, and the dynamical mergers of stellar binary black holes \citep[e.g.,][]{1987ApJ...321..199Q,Oleary09,2015MNRAS.448..754H, Antonini12, Wen03,2018ApJ...856..140H,2019MNRAS.488...47F,Paper1, Paper2, 2018CmPhy...1...53C,2023PhRvD.107d3009X}.

In order to precisely reveal the characteristics of these events, it is of fundamental importance to understand the stellar dynamics that drive the evolution of the orbits of stars and compact remnants in the vicinity of MBHs.

The stellar dynamics around the MBHs are driven by a number of different relaxation processes, e.g., the two-body relaxation, resonant relaxation (RR) \citep{RT96,2006ApJ...645L.133H}, and other dynamical effects, e.g., the tidal field of the MBH, relativistic orbital precession, gravitational wave orbital decay.

In the case of multiple mass components, mass segregation can drive the lighter components outward and the heavier ones migrate in, which is important for a number of investigations of the properties of the cluster, such as density profiles \citep[e.g.,][]{1977ApJ...216..883B,Alexander09,2010ApJ...708L..42P} and rates of EMRIs \citep[e.g.,][]{2006ApJ...645L.133H,2011CQGra..28i4017A}. If the particles are binaries, the dynamics can be even more complex, e.g., the ionization of binaries, exchanges of binary components, Kozai-Lidov effects \citep[e.g.,][]{2009ApJ...700.1933H, Paper1}. These complicated effects, which cover from small to large scales of the cluster, impose difficulty in obtaining accurate pictures of the evolution and outcomes of particles of different types and masses near the MBH.

The two-body relaxation process, which is essential for stellar dynamics near the MBH, has already been significantly investigated over the past several decades. Most of them adopt the one-dimensional Fokker-Planck (FP) analytical method (hereafter 1D-FP method), which mainly considers only the evolution in energy under the assumption of an isotropic distribution of angular momentum \citep{1977ApJ...211..244L, BW76, 1977ApJ...216..883B,2013ApJ...764...52B, 2005ApJ...629..362H,Alexander09,2009ApJ...698L..64K}.

As the relaxation on angular momentum is usually much faster than that in energies (unless in a near-circular orbit), its evolution can be considered approximately independent, and the effects of the loss cone can be approximately included in the evolution of energy \citep{1977ApJ...211..244L,1977ApJ...216..883B}. Usually, the orbits of particles can be considered approximately Keplerian near the MBH, and the effects of the stellar potential are important only at the outer parts of the cluster (i.e., outside the influence radius of the MBH) \citep{1979ApJ...234..317M}.

The numerical methods that consider both the evolution of energy and angular momentum in the FP equations have also been developed, either by Monte-Carlo methods \citep{SM78,Baror16}, finite differential methods \citep{1978ApJ...226.1087C,1995PASJ...47..561T}, or finite element methods \citep{1995PASJ...47..561T}. However, these previous studies are limited to only one mass component (assuming equal masses for all particles) and thus cannot handle the process of mass segregation if there are multiple mass components in the cluster. \citet{1997PASJ...49..547T} extends the finite differential/element methods to include multiple mass components, although we notice that it is difficult to expand further on the method by including other dynamical effects, such as stellar collisions and those of the binaries.

Another approach is to use direct N-body numerical simulations \citep[e.g.,][]{2014ApJ...794..106A,2010PhRvD..81f2002M,2004ApJ...613L.109P,2019MNRAS.484.3279P,2004ApJ...613.1133B}, most of which adopt the N-body code series \citep{1999PASP..111.1333A}. However, it is usually expensive to integrate the orbits of stars for more than several relaxation timescales, and sometimes it is necessary to use specialized hardware such as GRAPE \citep{2004ApJ...613L.109P,2004ApJ...613.1133B, 2014ApJ...794..106A}. Limited by the resolution and the number of particles in the simulation, it is challenging to simulate scales below $10^{-3}\sim10^{-2}$ of the gravitational influence radius of the MBH \citep{2019MNRAS.484.3279P}. Also, the N-body simulations are difficult to include various processes that happen at small scales as mentioned earlier.

Another important approach is the Monte-Carlo method based on the H'{e}non scheme \citep{1961AnAp...24..369H,2001A&A...375..711F}. The relaxation in such a method can be considered by shell-like particles, and the density of particles is updated after each time step. The method is flexible to include various dynamical effects and multiple mass components, which has already been applied for simulations of globular clusters \citep[e.g.,][]{2000ApJ...540..969J} and galactic nuclei \citep{2001A&A...375..711F,2002A&A...394..345F, 2006ApJ...649...91F}.

A number of interesting dynamical phenomena, e.g., tidal disruption events,
stellar collisions, and different kinds of gravitational wave sources such as
EMRIs, are results of particles evolving in both energy and angular momentum
and embedded in a cluster consisting of multiple mass components. Thus, it is
necessary to develop a Monte-Carlo method to simulate the evolution of
particles in such an environment with acceptable accuracies and numerical
costs. Here we develop a different Monte-Carlo approach (\GNC)\footnote{The
source code can be cloned in\\
\href{https://github.com/zhangfupeng-gzhu/GNC.git}{https://github.com/zhangfupeng-gzhu/GNC.git}}
based on \citet{SM78}, which is extended such that, for the first time in the
literature, it can obtain the dynamical evolutions of objects with multiple
masses based on two-dimensional (both energy and angular momentum) FP
equations. The method is also an overhaul of the previous version of the code
in \citep{Paper1,Paper2}. One of the advantages of \GNC{} is that it is very
flexible in including various kinds of stellar objects and different dynamical
effects mentioned above. \GNC{} can obtain statistically accurate results for
rare objects by varying the weightings of particles. Our code can be applied to
a number of important dynamical phenomena in galactic nuclei in the future,
providing a more consistent picture of dynamics for particles of various types
and across a spectrum of masses.

This paper is organized as follows. In Section 2, we introduce the core method of \GNC, including the Monte-Carlo method of the two-body relaxation, the boundary conditions, the weighting of particles, and the general Monte-Carlo routines in obtaining steady-state solutions. In Section 3, we compare our simulation results with those of previous studies for models consisting of one or multiple mass components. We test the relaxation processes, the density profiles, and the consumption rates of particles in the loss cone. Discussion and conclusions of the work will be presented in Section 4.

\section{The method}
\GNC{} is a Monte Carlo code that can calculate the evolution of particles in energy and angular momentum spaces when a central massive black hole (MBH) is present. In our previous works~\citet{Paper1} and~\citet{Paper2}, we have considered the evolution of binary black holes in a fixed background of single stars or black holes. In this work, we have made many improvements to \GNC{} to transform it into a general tool for obtaining steady-state solutions of multiple mass components. It is important to note that we only provide a description of the basic version of \GNC{} in this context, and we do not incorporate other dynamical effects (e.g., gravitational wave orbital dissipation) apart from the two-body relaxation and Newtonian loss cone. Further details of the method are presented in the following sections.

\subsection{Two body FP evolution of multiple mass components}

Per definition, the MBH dominates the dynamics of the cluster within the gravitational influence radius $r_h$. According to the $\bh-\sigma$ relation from~\citet{2013ARA&A..51..511K},
\be
\sigma_h=200~{\kms}\left(\frac{\bh}{3.16\times10^8\msun}\right)^{\frac{1}{4.42}},
\ee
where $\sigma_h$ is the velocity dispersion of the galaxy, 
the gravitational influence radius $r_h$ is given by, 
\be
r_h=\frac{G\bh}{\sigma_h^2}=r_0\left(\frac{\bh}{4\times10^6\msun}\right)^{0.55},
\label{eq:rh}
\ee
where $r_0=3.1~{\rm pc}$ is the influence radius for the MBH in Milky Way. 

This radius is roughly consistent with the breaking radius measured about $\sim 3$pc in our Galactic center
~\citep{2018A&A...609A..27S}. Then the density at $r_h$, i.e., $n_h$, is given by 
\be
n_h\propto\frac{\bh}{r_h^3}=n_0 \left(\frac{\bh}{4\times10^6\msun}\right)^{-0.65}
=n_0 \left(\frac{r_h}{r_0}\right)^{-1.18},
\label{eq:nh}
\ee
where $n_0$ is the density of stars at $r_0$ for Milky Way. 
Galactic center observation suggest that $n_0=3\times10^4$ pc$^{-3}$  at distance of $r_h=3.1$pc~\citep{2018A&A...609A..27S} 
and $n_0=2\times10^4 $ pc$^{-3}$~\citep{2003ApJ...594..812G} for $\bh=4\times10^6\msun$.
Here we adopt the latter value of $n_0$, for simplicity. 

If the cluster contains multiple mass components, the two body relaxation on energy $E$ and 
angular momentum $J$ of each of the components follows the FP differential equations that is coupled with those 
of other components. Here we define $E=G\bh /(2a_2)$ and $J=\sqrt{G\bh a_2(1-e_2^2)}$, 
where $a_2$ and $e_2$ are the semimajor axis (sma) and 
eccentricity of the outer orbit of the particle circling MBH. 
For the $\alpha$-th component, $\alpha=1,\cdots, N_m$, where $N_m$ is the total number of mass components, it is destribed by 
\be\ba
\frac{\pl N_\alpha(E,J)}{\pl t}&=-\frac{\pl [N_\alpha(E,J)D^E_{\alpha}]}{\pl E}-\frac{\pl [N_\alpha(E,J)D^{J}_{\alpha}]}{\pl J}\\
+&\frac{1}{2}\frac{\pl^2 [N_\alpha(E,J)D^{EE}_{\alpha}]}{\pl E^2}+\frac{1}{2}\frac{\pl^2 [N_\alpha(E,J)D^{JJ}_{\alpha}]}{\pl J^2}\\
+&\frac{\pl^2 [N_\alpha(E,J)D^{EJ}_{\alpha}]}{\pl E\pl J},
\label{eq:EJequation}
\ea
\ee
where $N_\alpha(E,J)$ is the number distributions of particles of the $\alpha$-th mass component in $E-J$ spaces. 

The diffusion coefficients $D^E, D^J$ describe the drift, and $D^{EE}$ and $D^{JJ}$ describe the 
scatterings of $E$ and $J$. $D^{EJ}$ describe the correlation of scatterings between $E$ and $J$. 
These diffusion coefficients are coupled with the distribution functions in other mass bins. 
Specifically, for the $\alpha$-th component with mass $m_\alpha$, the drift terms  are
given by
\be\ba
\frac{D^E_\alpha}{E}&=\sum_\beta m_\beta^2 \left(\frac{m_\alpha}{m_\beta} \Gamma^{110}_\beta -\Gamma_\beta^0\right)\\
\frac{D^J_\alpha}{J_c}&=\frac{1}{j} \sum_\beta m_\beta^2 \left(\Gamma^{310}_\beta-\frac{1}{3}\Gamma_\beta^{330}\right.\\
&\left.-\frac{1}{2}\frac{m_\alpha+m_\beta}{m_\beta}j^2\Gamma_\beta^{111}+\frac{5-3j^2}{12}\Gamma_\beta^0\right),
\label{eq:dedj1}
\ea\ee
the diffuse and cross terms are given by
\be\ba
\frac{D^{EE}_\alpha}{E^2}&=\frac{4}{3} \sum_\beta m_\beta^2 \left(\Gamma_\beta^{13-1} +\Gamma_\beta^0\right)\\
\frac{D^{JJ}_\alpha}{J^2_c}&=\sum_\beta m_\beta^2 
\left[\frac{j^2}{2} \Gamma_\beta^{131}+ 2\Gamma_\beta^{310}\right.\\
&\left.-\frac{2}{3}\Gamma_\beta^{330}-\frac{j^2}{2}\Gamma^{111}_\beta+\frac{5-3j^2}{6}\Gamma_\beta^0\right]\\
\frac{D^{EJ}_\alpha}{J_cE}&=-\frac{2 j}{3}\sum_\beta m_\beta^2 \left(\Gamma_\beta^{130}+\Gamma_\beta^0\right),
\label{eq:dedj2}
\ea\ee
where $\beta$ runs over all mass components. $j=J/J_c$ is the dimensionless angular momentum and 
$$J_c=\frac{G\bh}{(2E)^{1/2}}$$ is the angular momentum for circular orbits.
$\Gamma_\beta$ functions can be found in~\citet{Baror16} or \citet{Paper1}, 
which depends on the dimensionless function $\bar g_\beta(E)$ of the $\beta$-th component, and it will be explained later. 

The relation between the number density $N_\alpha(E,J)$ and
 the distribution function $f_\alpha(E,J)$ in phase spaces is given by 
\be\ba
N_\alpha(E,J)dEdJ&=\int_{\mathbf{V}} f_\alpha(E,J) d\mathbf V d\mathbf v\\
&=8\pi^2Jf_\alpha(E,J)P(E)dEdJ,
\ea\ee
where $P(E)=2\pi \bh/(2E)^{3/2}$ is the orbital period around a MBH with mass $\bh$, $\mathbf V$ and $\mathbf v$ is the 
3-dimensional position and velocity, respectively. 
When discussing the dynamics and distributions of particles,
it is more convenient to adopt the following dimensionless form:
\be\ba
g_\alpha(x,j)&=(2\pi\sigma_\star^2)^{3/2}n_\star^{-1}f_\alpha(E,J) \\
x&=\frac{E}{\sigma_\star^2};~j=\frac{J}{J_c},
\label{eq:gxj_fEJ}
\ea\ee
where $\sigma_\star=\sigma_h$, $n_\star=n_h$ is the velocity dispersion and 
density of the reference star at $r_h$, respectively. 

The dimensionless number distribution $N_\alpha(x,j)$
in the $\alpha$-th mass bin is then given by 
$N_\alpha(E,J)=N_\alpha(x,j)\sigma_\star^{-1}\sqrt{2}x^{1/2}(G\bh)^{-1}$. 
According to Equation~\ref{eq:gxj_fEJ}, we have 
\be
g_\alpha(x,j)=N_\alpha (x,j)\pi^{-3/2}n_\star^{-1}\sigma_\star^6x^{5/2}j^{-1}(G\bh)^{-3}.
\label{eq:galpha}
\ee
We can now define $\bar g_\alpha(x)$, the $j$-averaged dimensionless parameter by 
\be
\bar g_\alpha(x)=2\int_0^1 g_\alpha(x,j) j dj.
\label{eq:bar_galpha}
\ee
Then $\bar g_\alpha(x)$ can then be used for calculations of the $\Gamma_\alpha$ functions in the corresponding mass bin.

\subsection{Boundary conditions}
\label{subsec:boundary_condition}
\begin{figure*}
\center
\includegraphics[scale=0.75]{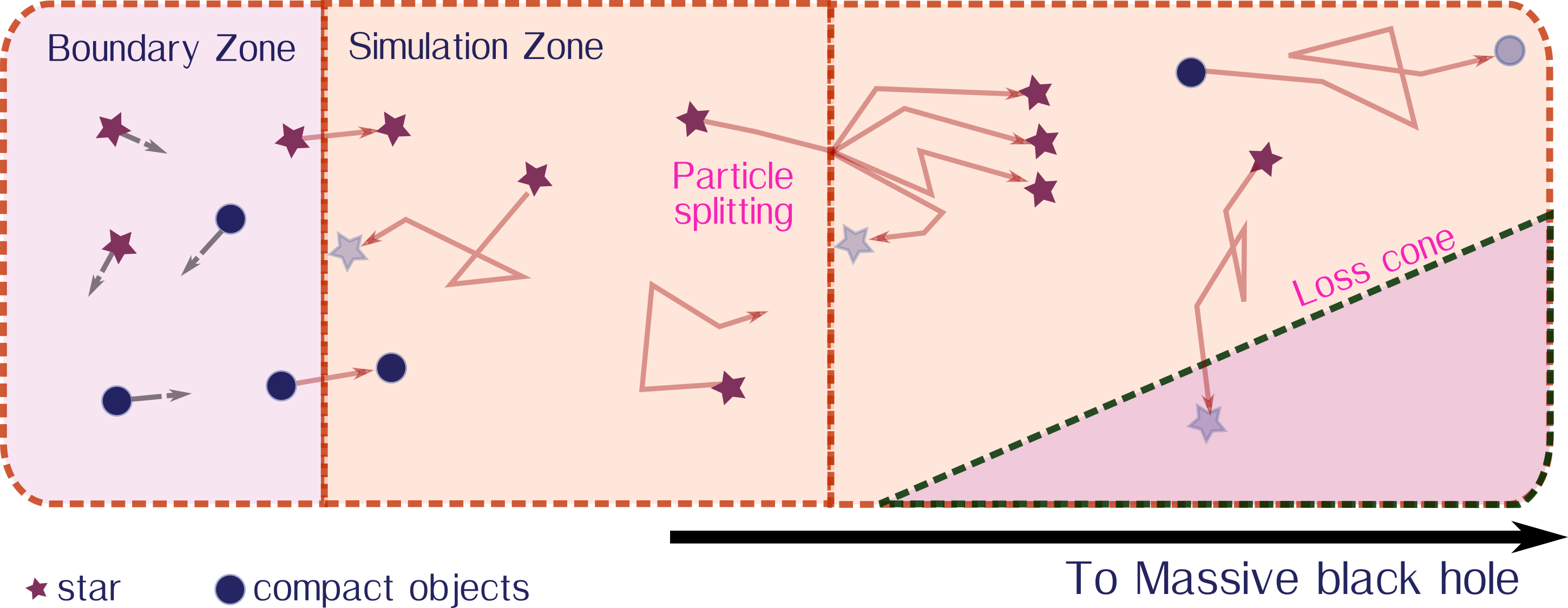}
\caption{This illustration shows the treatment of boundary and the evolution of multiple types of particles. 
Particles outside the outer boundary (the left panel) flow into the cluster 
following the procedure described in Section~\ref{subsec:boundary_condition} and 
Appendix~\ref{apx:MC_method}. 
Inside the outer boundary particles move in the two dimensional $E-J$ space (mid and right panel). 
Particles pass through $x=10^i$, $i=1,\cdots,4$ from the left will 
be split to multiple clone particles, of which have smaller weights to ensure 
the conservation of particle number (the right panel). The boundary zone ranges from 
$x=0.03$ to $x_B$, where $x_B=0.05$ and the simulation zone ranges 
from $x=x_B$ to $x_{\rm max}=10^5$. Symbols filled with light colors show the possible 
positions where the simulation of a particle stops.
 }
\label{fig:drawing}
\end{figure*}

Particles will be destroyed if they approach too close to the MBH. Therefore, it is natural to impose a vanishing boundary condition in the inner part of the cluster. In this case, we set $x_{\rm max}=10^5$ as the inner boundary, which corresponds to a distance of approximately $\sim3.2$ AU for an MBH with $\bh=4\times10^6\msun$. 

On the other hand, if a particle moves too far away from the influence radius, it may become unbound from the MBH. Ideally, the boundary should be set at $x=0$. However, approaching this value numerically is not feasible, so we set the outer boundary at $x=x_B=0.05$, which is smaller than the value used by~\citet{SM78} (they set $x_B=0.2$). Particles with energies larger than this value are considered unbound from the MBH. The unbound stellar populations should be asymptotically similar to those of the bulge stars. Therefore, it is reasonable to set a fixed given number density (or number ratio) for different particle types, which corresponds to the Dirichlet boundary condition at $x_B$ for Equation~\ref{eq:EJequation}.

Suppose there are multiple mass components for $x<0$ (the unbound population), denoted by $m_\alpha=1,\cdots, N_m$, and each of them consists of $\beta=1,\cdots, N_{t\alpha}$ different types of particles. For example, we can set $\beta=1$ for stars, $2$ for stellar-mass black holes (SBHs), $3$ for neutron stars (NSs), $4$ for white dwarfs (WDs), and $5$ for brown dwarfs (BDs). Let $s_\beta(m_\alpha)$ represent the number density ratio of the $\alpha$-th mass component of type $\beta$ relative to the total number of stars at the outer boundary. For stars, we have $\int s_\beta(m_\star) dm_\beta=1$. If we assume that all stars have equal masses, then we simply have $s_\beta(m_\star)=1$.

Assume violent relaxation boundary conditions, the velocity dispersion of all components is the 
same as those of the reference star~\citep{Alexander09}.
For the objects with type $\beta$ in the
  $\alpha$-th mass bin, the distribution function 
$g_{\alpha\beta}(x,j)$, $x\le x_B$, at the boundary is given by

\be\ba
g_{\alpha\beta}(x<x_B,j)=s_\beta(m_\alpha)G(j),~ {\rm if~}x=x_B
\label{eq:gj}
\ea
\ee

%If the particles are in the full loss cone 
%region, we should have $N(E_B,J)\propto J$~\citep{1977ApJ...211..244L}, and 
%$N(E_B,J)\propto J\ln J/J_{\rm lc}$ for empty loss cone region
%(For the definition of full and empty loss cone see Section~\ref{subsec:analy_method}). 
As we will discuss later (in Section~\ref{subsec:flux}), when the mass of the MBH is around $\bh=10^4-10^8\msun$, the particles, especially compact objects, slightly inside the outer boundary, follow a form closer to the full loss cone (See also Equation~\ref{eq:gj_fulliso}). However, if the mass of the MBH is larger than $10^8\msun$, the distribution gradually transitions to the empty loss cone condition (See also Equation~\ref{eq:gj_empiso}). Since we primarily focus on galaxies containing $\bh<10^8\msun$, which are more commonly found in the nearby universe, we simply assume a full loss cone condition for all simulations conducted in our study. Therefore, we set\footnote{As $N(x,j)\propto j$, there are very few samples with small values of $j$. However, these samples can cause significant Monte-Carlo fluctuations in the model results. To avoid this, we set a lower limit of $j>0.0044$ (or $e<0.99999$) when generating samples at the outer boundary.}
$
G(j)=1.
%\label{eq:gj_iso}
$

For the isotropic boundary condition, we follow the approach of~\citet{SM78} and assume that the unbound population ($x<0$) consists of bulge stars with isothermal distributions and a Maxwellian velocity distribution. Thus, we have $g_{\alpha\beta}(x<x_B,j)=s_\beta(m_\alpha)e^{x}$ for $x<0$. This isotropic outer boundary is consistent with observations of our Galactic center~\citep{2017MNRAS.466.4040F}.

With these boundary conditions, \GNC{} will generate a continuous flow of particles into the cluster. Further details about the Monte Carlo methods can be found in Appendix~\ref{apx:MC_method}. Figure~\ref{fig:drawing} provides an illustration of the boundary methods.

\subsection{Particle Weights and Particle Splitting Method}
\label{subsec:weighting}

%The clone scheme and the weighting of particles are now improved such that 
%the code can easily handle the interplay and possible exchanges between particles of different types 
%and different masses. The method in determining the weights of particles, compared to the previous version of \GNC, 
%is adjusted such that the rare particles can have better statistics.

We acknowledge that there is a difficulty in the Monte Carlo simulation due to the significant variation in the number of particles in different mass bins. For instance, in a two-component model, the number ratio of single black holes ($\sim 10-40\msun$) to reference stars ($\sim 1\msun$) is on the order of $10^{-3}$~\citep{2006ApJ...645L.133H} or even smaller. Therefore, if each particle has the same weight, the number of stars will dominate over the number of single black holes. To increase the efficiency of the simulation and improve the statistics of rare particles, it is more convenient to assign different weights to particles in different mass bins. We denote this weighting factor as $w_\alpha$ for the $\alpha$-th mass bin. For example, in the case of $\bh=4\times10^6\msun$, we can set $w_1=400$ for stars (or $w_2=10$ for single black holes), meaning that one particle of stars (or single black holes) in the simulation represents 400 (or 10) particles. In this way, we find that in the steady state of the simulation, the total number of single black holes is comparable to the number of stars. For particles in other mass bins and for different MBH masses, we can adjust the value of $w_\alpha$ to ensure an adequate number of particles for accurate statistics.

Previously, \citet{SM78} used a particle cloning scheme to increase the statistics of particles in the inner regions. The previous version of \GNC{} also adopted this scheme~\citep{Paper1}. However, this scheme requires labeling particles as "original" or "clones," which increases the complexity for future expansions of the code. For example, when a "clone" star merges with an "original" star after a stellar collision, or when a "clone" single black hole becomes a component of an "original" binary system after a three-body exchange, it becomes challenging to determine whether the merged/exchanged product should be classified as an "original" particle or a "clone" particle.

To address this issue, we have improved the cloning scheme to a more general splitting method. When a particle moves inward and crosses the energy $x/x_0=10^i$, the particle splits into $\Pi$ clones (or creates $\Pi-1$ clones), where $i=0,\cdots, 4$ and $x_0$ is an arbitrarily dimensionless energy above which the clone scheme is implimented. Conversely, when any particle moves upward and crosses back to those energies (regardless of whether it is a clone or an original particle), there is a chance of $1-1/\Pi$ that it will be deleted from the simulation. To obtain correct statistical results, we need to correct the weight of the split particle by another factor, which is given by $w_c=1/\Pi^{\rm L}$, where $L={\rm max}[{\rm Int}(\log_{10} (x/x_0)+1),0]$. The particle splitting method is also illustrated in Figure~\ref{fig:drawing}. We note that if $\Pi$ is the same for particles in different mass bins, there will be an overabundance of massive particles concentrated in the inner parts of the cluster due to mass segregation. To address this issue, we can assign different values of $\Pi$ for each mass bin and decrease $\Pi$ to a lower value for mass bins with larger masses. In the models presented in this work, typically $\Pi$ varies from $30$ to $4-10$ as the masses increase from $\lesssim 1\msun$ to $\ga 10-40\msun$.

Suppose the ``uncorrected'' dimensionless distribution function at the outer boundary for an arbitrary mass bin $\alpha$ is given by $\bar g_\alpha^u(x_B)$. It needs to be corrected such that:

\be
\bar g_\alpha^u(x_B)w_n = \bar g_\alpha(x_B)=s(m_\alpha)=\sum_\beta s_\beta(m_\alpha).
\label{eq:galpha_norm}
\ee 

This correction determines a constant $w_n=s(m_\alpha)/\bar g^u_\alpha(x_B)$, which can be used for the correction to obtain the true weights of all particles. It is important to note that we can select an arbitrary mass bin for this correction, as the number ratio of particles beyond the outer boundary, $s_\beta(m_\alpha)$, remains unchanged during the simulation. Therefore, the value of $w_n$ obtained from the $\alpha$ mass bin is the same for other mass bins as well.

Finally, the true weights of any particle are determined as:
%
%Finally, the true weights of any particle is determined 
\be
w=w_n w_c w_\alpha
\ee

Normalization is usually necessary before the 3rd or 4th iteration, after which the weighting $w_n$ remains constant as the distribution near $x_B$ converges. It is important to note that, according to the normalization method and Equation~\ref{eq:galpha}, the true weight of particles corresponds to the objects in reality, as well as the associated results (e.g., event rates of tidal disruptions for stars), will be proportional to $n_0 r_0^3$, where $r_0$ and $n_0$ are given by Equations~\ref{eq:rh} and \ref{eq:nh}, respectively. This should be kept in mind when comparing our results with those from other authors.

\subsection{Steady-state solution of distribution, density and anisotropy}

The Monte Carlo simulation begins with an initial guess of distributions for different mass components, denoted as $N_\alpha(E,J)$. All particles undergo a simulation duration of $\delta \tau= 0.005\sim0.01T_{\rm rlx}(r_h)$, where $T_{\rm rlx}(r_h)$ is the 
two-body relaxation time at $r_h$. In the case of multiple components, it is given by~\citep{Binney87}:

\be
T_{\rm rlx}(r_h)=\frac{0.34 \sigma_h^3}{G^2\sum_\alpha m_\alpha^2 n_\alpha(r_h)\ln \Lambda},
\label{eq:trlx}
\ee
where approximately $\Lambda\simeq\bh/m_\star$.
After that, the distribution of particles $\bar g(x)$ is updated to a new one, which is then used to calculate the diffusion coefficients for the next iteration. 
Repeat this process for a sufficient number of times, and the distribution will gradually converge to a steady-state solution.
More details are shown in Appendix~\ref{apx:MC_steady}.

At any snapshot in time, the number density of particles in the mass bin $\alpha$ at a distance of $r$ is then given by
\footnote{Here we only consider the density of bound components. Including the unbound population, if we ignore the potential of stars, the density profile for $r\gtrsim r_h$ will be flattened to $\propto r^{-1/2}$~\citep{1978ApJ...226.1087C}, which deviates significantly from the profiles of $\propto r^{-2}$ expected from an isothermal cluster or those observed in the Galactic Center~\citep{2003ApJ...594..812G}. A better solution would be to specify detailed models of the outer parts of the cluster~\citep{BW76, 1979ApJ...234..317M}, with or without the stellar potentials. However, we notice that the models adopted by these studies are specified for globular clusters, not those of galactic nuclei.}

\be\ba
n_\alpha(r)&=\int^{\frac{\bh G}{r}}_{0}\int_0^{r\sqrt{2(\frac{\bh G}{r}-E)}} f_\alpha(E,J) \frac{4\pi JdJdE}{r^2v_r}
\label{eq:density}%\\
%&\simeq 2\pi^{1/2} n_h\int^{r_h/r}  \bar g_\alpha(x) \sqrt{\frac{r_h}{r}-x} dx
\ea\ee

We can also estimate the anisotropy at position $r$ by the parameter%~\citep{1997PASJ...49..547T}
\be
\mathscr{A}(r)=1-\frac{\langle v_t^2(r)\rangle }{2\langle v_r^2(r)\rangle}
\ee
where 
\be
\ba
\langle v_t^2\rangle=\frac{1}{n(r)}\iint v_t^2 f(E,J)  \frac{4\pi JdJdE}{r^2v_r}\\
\ea
\ee
and 
\be
\ba
\langle v_r^2\rangle=\frac{1}{n(r)}\iint v_r^2 f(E,J) \frac{4\pi JdJdE}{r^2v_r}
\ea\ee
is the velocity dispersion in tangential and 
radial directions, respectively. These values can be estimated 
according to $v^2_t=J^2/r^2$, $v^2_r=v^2-v_t^2$ and 
$v^2=2(\bh/r-E)$ for each particle.

\section{Simulations and results}
To test the results from \GNC, we need to compare them to previous studies, particularly those that have 
used one-dimensional Fokker-Planck (1D-FP) methods. The 1D-FP method assumes an isotropic distribution of angular momentum and considers only one-dimensional energy relaxations. This method has been extensively tested in the past and has shown general consistency with results obtained from other methods. Therefore, our main focus here is to compare our results with those obtained using the 1D-FP method, as it serves as a simplified version of our approach.
We provide a review of the relaxation process and the loss cone effect of the
 1D-FP method in Appendix~\ref{subsec:analy_method}.

\subsection{The Relaxation and density profile of nuclear star cluster}
\begin{figure*}
	\center
	\includegraphics[scale=0.6]{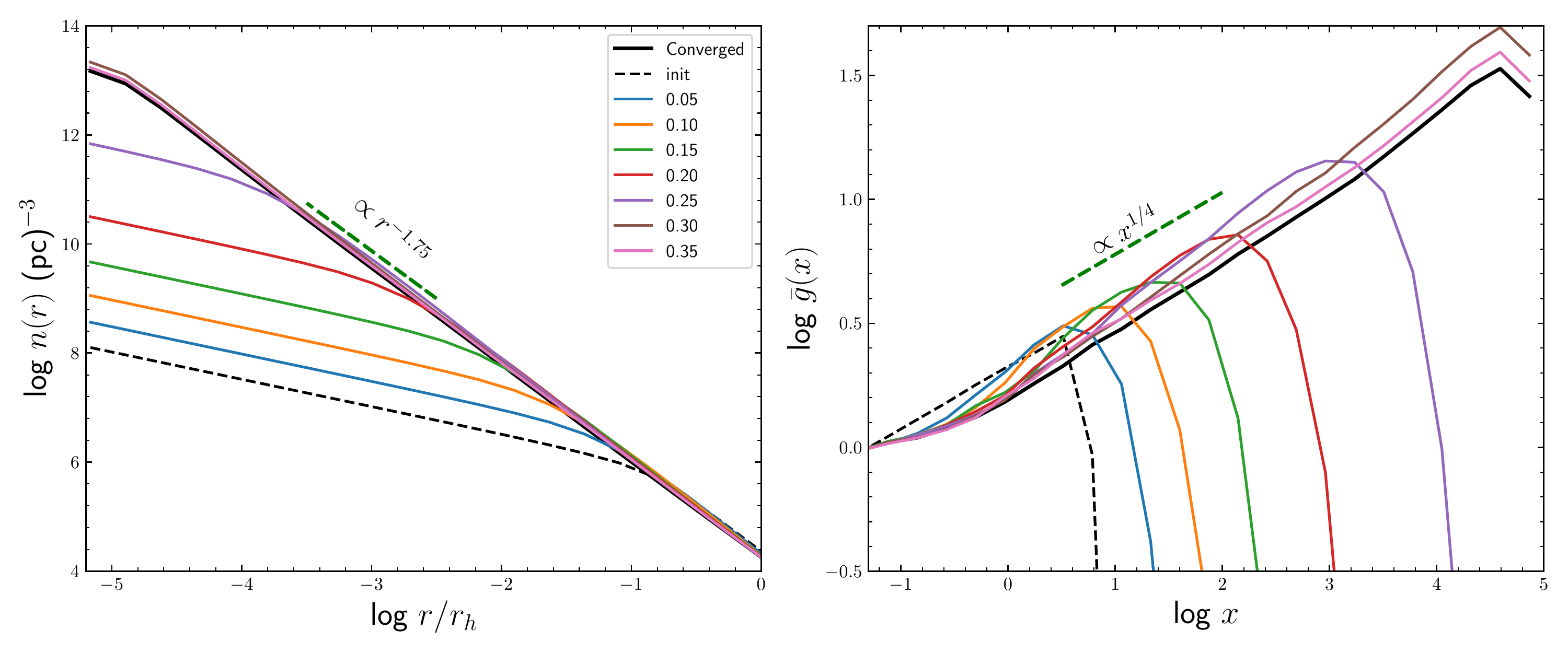}
	\caption{The evolution of distributions of equal mass star cluster ($m_\star=1\msun$, model M1) as  a
	function of time. Loss cone effects are ignored. Left panel: Density profile evolution of a cluster as a function of 
	time in unit of $T_{\rm rlx}$, where $T_{\rm rlx}$ is the two-body relaxation time at $r_h$.  
	The mass of the MBH is $\bh=4\times10^6\msun$. Right panel: Same as the left one but for the dimensionless 
	distribution function $\bar g(x)$. 
		}
	\label{fig:relax}
\end{figure*}

\begin{figure*}
\center
\includegraphics[scale=0.47]{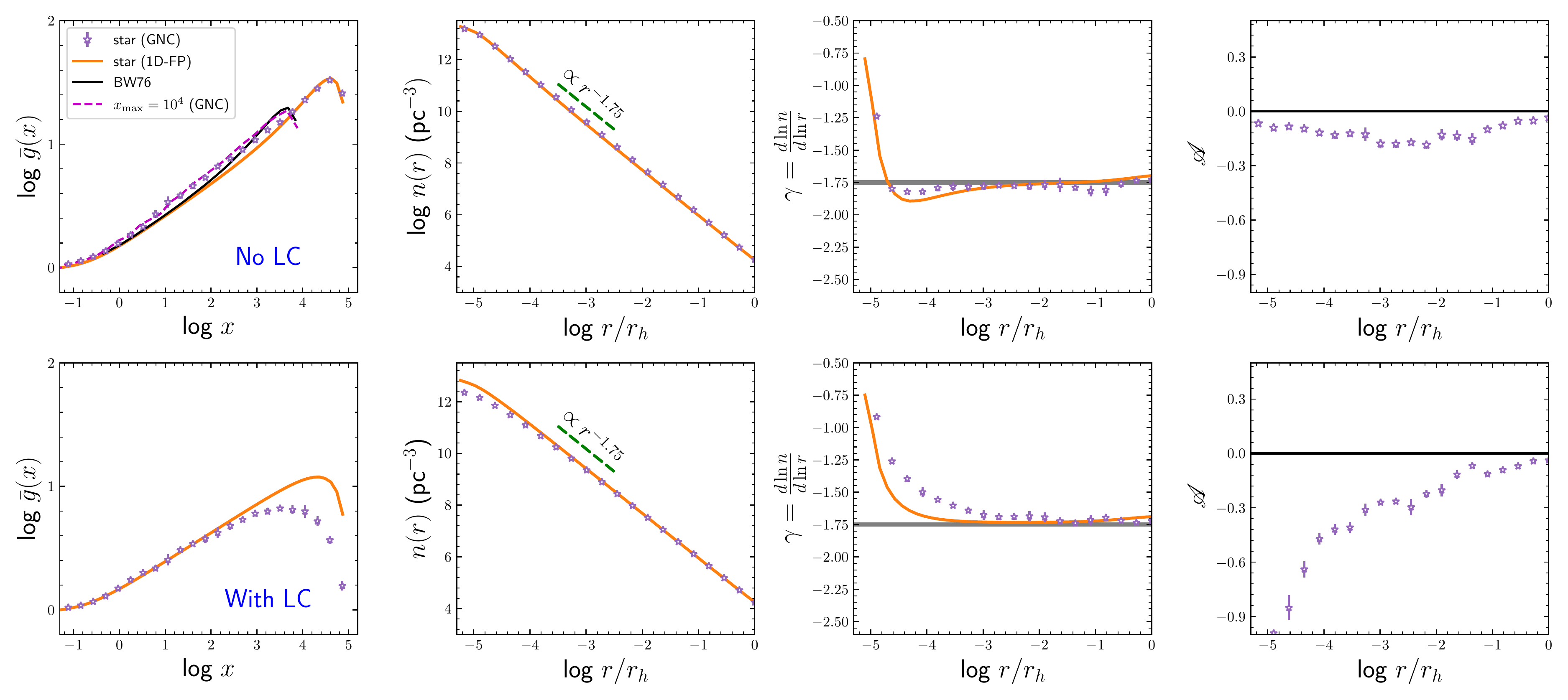}
\caption{Comparison of our method and those from 1D-FP methods for single mass of stars (model M1, See Table~\ref{tab:model}). 
Top panels and bottom panels are results ignoring and including the loss cone effect, respectively.
Panels from the left to the right are: 
The dimensionless energy distribution function $\bar g(x)$; 
the number density $n(r)$; The slope index of the number density $\gamma={\rm d} \ln n(r)/{\rm d} \ln r$; 
and the aniostropy. The colored lines in the first three panels are results from the 1D-FP method.
In the top left panel, the black solid line are the results from~\citet{BW76}, which adopt $x_{\rm max}=10^4$. 
The magenta dashed line are the results from our method adopting $x_{\rm max}=10^4$. 
%The black solid line in the second panel show the N-body simulation 
%results from~\citet{2004ApJ...613.1133B}. The black dashed line in the same panel show the Monte-Carlo 
%simulation result from~\citet{2006ApJ...649...91F}. Both of these two results are 
%renormalized according to our density at $r_h$. 
}.
\label{fig:fig_gE1}
\end{figure*}

\begin{figure*}
	\center
	\includegraphics[scale=0.7]{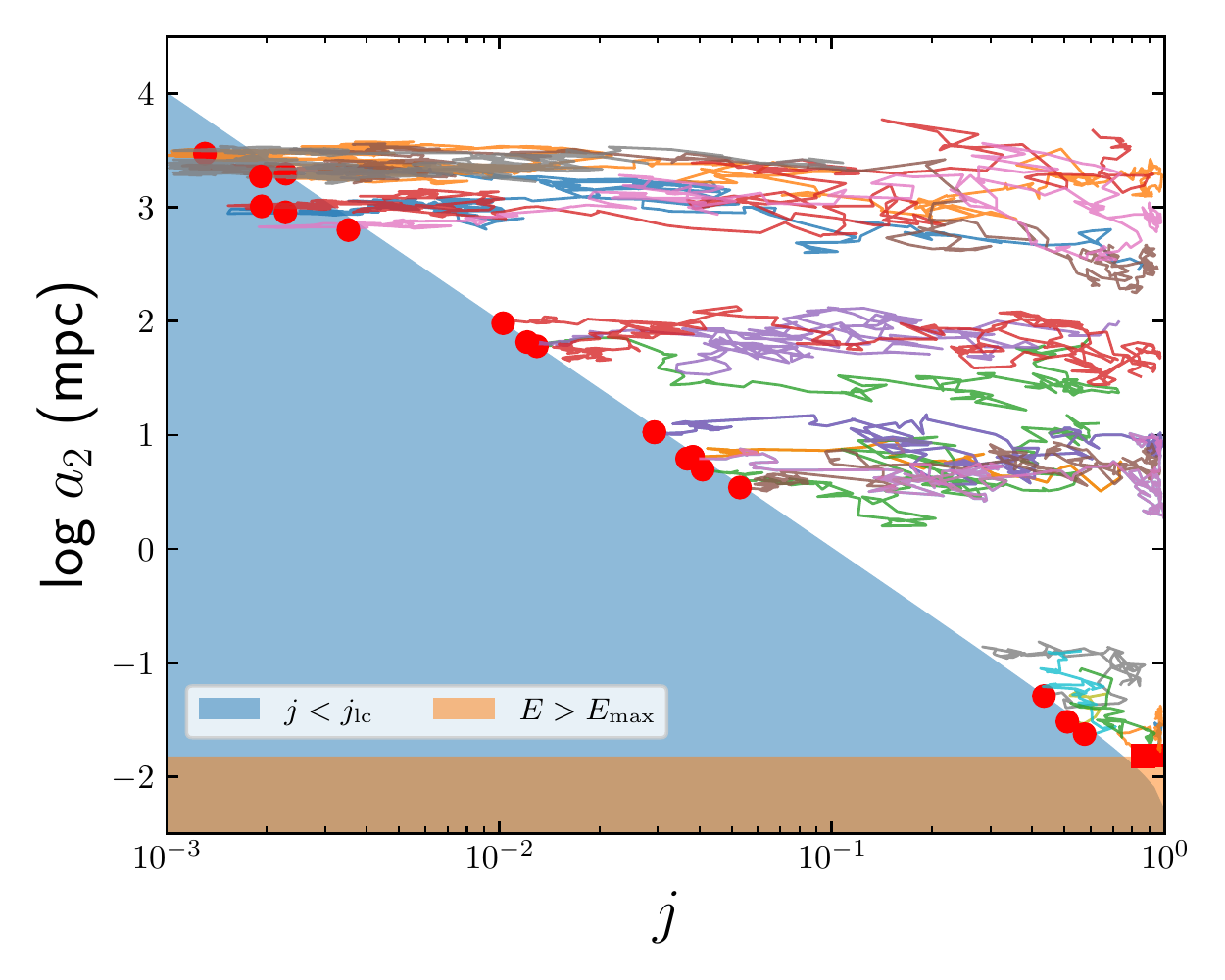}
	\includegraphics[scale=0.7]{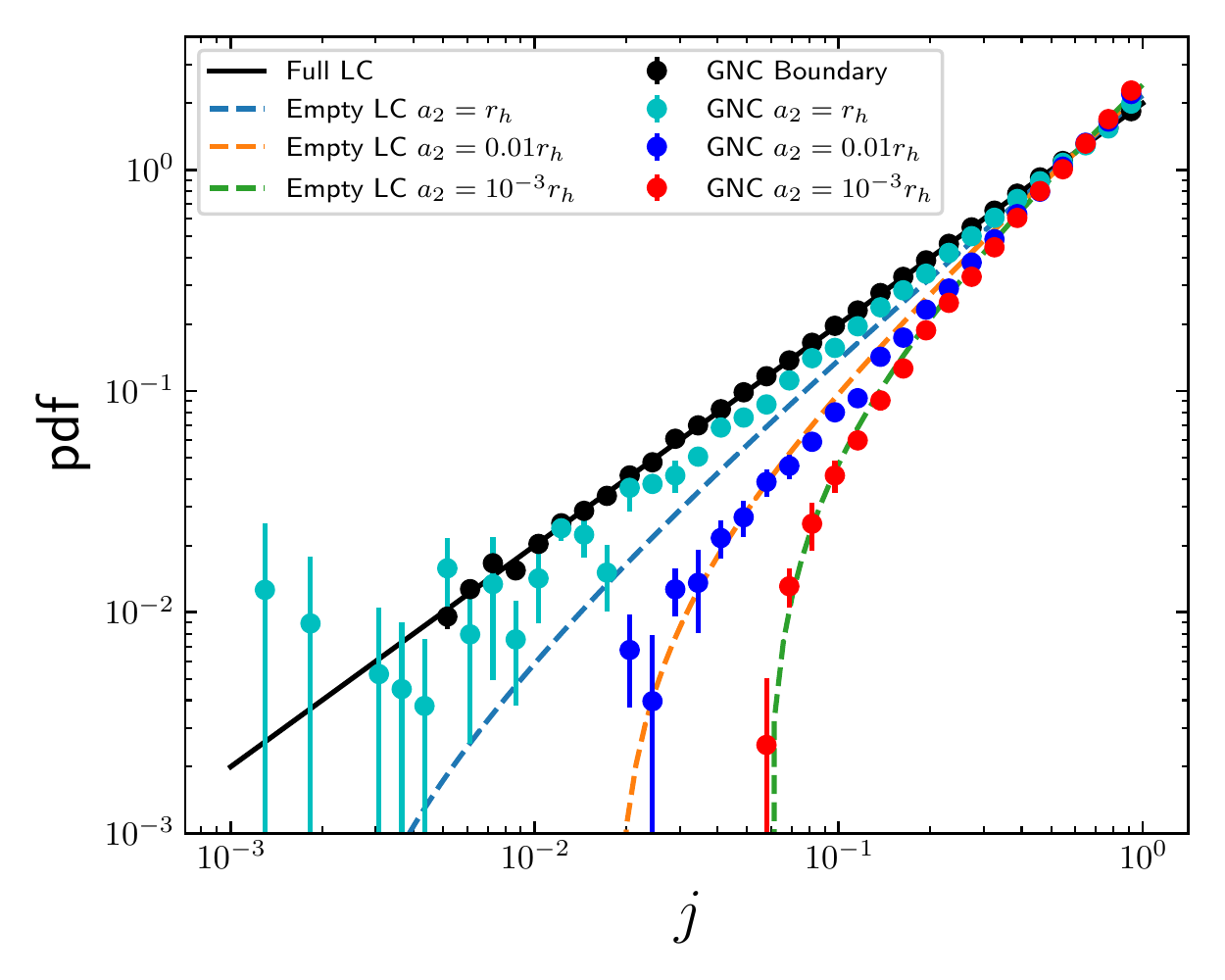}
	\caption{
 Left panel: Here are some examples of the trajectories of stars in \GNC{} around the central supermassive black hole (MBH) in model M1. The mass of the MBH is $\bh=4\times10^6\msun$, and all stars have an equal mass of $m_\star=1\msun$. The inner boundary of energy is $E_{\rm max}/\sigma^2_h=10^5$, which corresponds to approximately $r_h/(2\times10^5)\simeq 0.015 {\rm~mpc}\simeq 3.2$ AU of a circular orbit. The lines ending with red filled circles represent the trajectories of stars depleted by the loss cone, while those ending with filled squares represent trajectories that move inwards towards the inner boundary.
Right panel: The plot shows the normalized probability distribution function of the dimensionless angular momentum for particles with different semi-major axes. In the empty loss cone region, the distribution follows Equation~\ref{eq:gj_empiso}, while in the full loss cone region, it follows a distribution proportional to $j$. The particles at the boundary are located at $0.03<x<x_B$.
	}
	\label{fig:track}
\end{figure*}

\begin{table*}
\caption{Boundary conditions for test runs}
\centering
\begin{tabular}{cccc|cccc}\hline
 Model& Component& $m$ ($\msun$) & $s_\beta(m_\alpha)$ & Model
& Component& $m$ ($\msun$) & $s_\beta(m_\alpha)$ \\
\hline
M1 & star & $1$  &  $1$    & \multirow{5}{*}{M5} & star & $1$           &  $1$          \\ 
\cline{1-4}
\multirow{2}{*}{M2} & star & $1$           &  $1$      & & BD   & $0.05$        &  $0.2$          \\
					& SBH   &  $10$        & $10^{-3}$  & & WD  & $0.6$          & $0.1$   \\
\cline{1-4}
\multirow{3}{*}{M3} & star & $1$           &  $1$       & & NS  & $1.4$          & $10^{-2}$   \\
					& NS  & $1.4$           & $10^{-2}$ & & SBH  & $10$          &  $10^{-3}$   \\
\cline{5-8}					
					& SBH  & $10$          &  $10^{-3}$  & & & &   \\
\cline{1-4}
\end{tabular}
\tablecomments{Boundary conditions of test models. 
}
\label{tab:model}
\end{table*}

To test whether \GNC{} accurately simulates the dynamical evolution of particles around the 
central supermassive black hole (MBH), we first examine the relaxation and density profile of the 
nuclear star cluster using various test models with one or multiple mass components. The boundary conditions 
for different test models, ranging from 1 to 5 components, are shown in Table~\ref{tab:model}. 
We also discuss and analyze the results for models with a spectrum of masses in this section. We begin with a MBH mass of $\bh=4\times10^6\msun$, which corresponds to the mass of the Milky Way black hole~\citep{2009ApJ...692.1075G}, but we vary the MBH mass later in this section.

\subsubsection{Single-mass component}
We first examine the results from model M1, which consists of stars with equal 
masses ($m_\star=1\msun$). The left and right panels of Figure~\ref{fig:relax} show the evolution of the density profile and the energy distribution at different times of the simulation, respectively. We observe that the convergence time is around $\sim0.4T_{\rm rlx}(r_h)$ for particles inside a radius of $0.1r_h$, where $T_{\rm rlx}(r_h)$ is the two-body relaxation time at $r_h$ given by Equation~\ref{eq:trlx}. This timescale is consistent with the results of \citet{2004ApJ...613L.109P}, who also used N-body simulations. It is evident that the relaxation of particles becomes faster as they get closer to the central MBH.

Figure~\ref{fig:fig_gE1} compare the results of M1 from \GNC{} and those from 1D-FP methods. 
The two methods are well consistent, for the energy distribution $\bar g(x)$, 
the density profile $n(r)$, and the cumulative number distribution $N(<r)$. 
The discrepancies of results between these two methods at the inner parts of the cluster 
are slightly larger if additionally the loss cone is considered. The difference is mostly apparent in $\bar g(x)$, as it 
will amplify the difference in density profile by a factor of $\propto x^{3/2}$.

As we assume isotropic boundary conditions, the anisotropy of the particles is small 
($\mathscr{A}\sim0$) at the outer parts of the cluster. However, it becomes slightly tangentially anisotropic ($\mathscr{A}\sim-0.2$) in the middle of the cluster, even without the presence of a loss cone (top right panel of Figure~\ref{fig:fig_gE1}). If the particles are additionally depleted by the loss cone, the tangential anisotropy is increased to $\mathscr{A}\lesssim-0.6$ for $r<10^{-3}r_h$ (bottom right panel of Figure~\ref{fig:fig_gE1}). This small anisotropy developed in the cluster may explain the slight differences between our method and the 1D-FP methods, as the latter assumes isotropic distributions.

The tangential anisotropy observed in the inner parts of the cluster is consistent 
with the findings of \citet{1978ApJ...226.1087C}, who used finite differential methods 
in two-dimensional $E-J$ spaces. Observations of the Galactic center also suggest that the nuclear star cluster exhibits isotropy at outer parts but becomes tangentially anisotropic at inner parts~\citep{2017MNRAS.466.4040F}.

The slope index of the density profile ($\gamma=d\ln~n(r)/d\ln r$) of 
stars in model M1 can be seen in the third panel of Figure~\ref{fig:fig_gE1}. We observe that without the loss cone, $\gamma\simeq-1.75$ for $10^{-4}\lesssim/r_h\lesssim 0.1r_h$, which is the expected value for a star cusp around a MBH~\citep{BW76,1977ApJ...211..244L}. The slope of stars in the inner parts of the cluster ($r<0.01r_h$) becomes slightly shallower ($\gamma=-1.6\sim-1.7$) when the loss cone effects are included, and the dimensionless distribution function $\bar g(x)$ drops quickly as the inner boundary is approached.

The left panel of Figure~\ref{fig:track} shows examples of star trajectories in the simulation of model M1 when the loss cone is considered. It is evident that in most parts of the cluster, stars evolve along elongated trajectories, allowing them to efficiently reach high eccentricities. This is primarily because the relaxation in angular momentum is usually much faster than the relaxation in energy, except for nearly circular orbits.

The dynamics can be divided into two regions: the full loss cone region with $x\ll x_c$ (or $a_2\gg a_c=\bh G/(2 x_c)$), as defined in Equation~\ref{eq:analy_qxc}, and the empty
loss cone region with $x\gg x_c$ (or $a_2\ll a_c$). For this specific case with $\bh=4\times10^6\msun$, we find that $x_c\simeq 0.9$, corresponding to $a_c\simeq0.6r_h\simeq1.7$ pc. Figure~\ref{fig:track} demonstrates that near the full loss cone region, orbits of stars can still evolve without being immediately disrupted by tidal forces. They are only consumed if they remain within the loss cone for a sufficient duration to pass through the pericenter. Some of these stars can even move out of the loss cone before being depleted. Stars that remain inside the empty loss cone are highly likely to be disrupted since, within an orbital period, their angular momentum changes are too small to remove them from the loss cone. In the inner part of the cluster, where the relaxation timescale is much shorter, some low-eccentricity stars can sink towards the inner boundary ($x>10^5$, or $a_2\lesssim10^{-2}$ mpc for $\bh=4\times10^6\msun$) without being disrupted.

In the empty loss cone region ($a_2\ll 0.6r_h$ for model M1), the distribution of angular momentum follows the expectation of Equation~\ref{eq:gj_empiso}. The right panel of Figure~\ref{fig:track} shows that for particles with $a_2=0.01r_h$ or $10^{-3}r_h$, the distribution $N(j)$ obtained from \GNC{} is in good agreement with the expected distribution. For particles with semi-major axes $a_2=r_h$, since $a_2\sim 2a_c$ rather than $a_2\gg a_c$, the distribution $N(j)$ follows the form of the full loss cone region ($N(j)\propto j$) more closely than that of the empty loss cone region.

All of these results, including the relaxation timescale, the energy and density profile evolution, and the $E-J$ evolution of particles, suggest that \GNC{} yields results that are consistent with expectations from the literature, particularly in the case of single-mass particles in galactic nuclei.

\subsubsection{Multiple mass components}
\label{subsec:multiple_masses}
\begin{figure*}
	\center
	\includegraphics[scale=0.6]{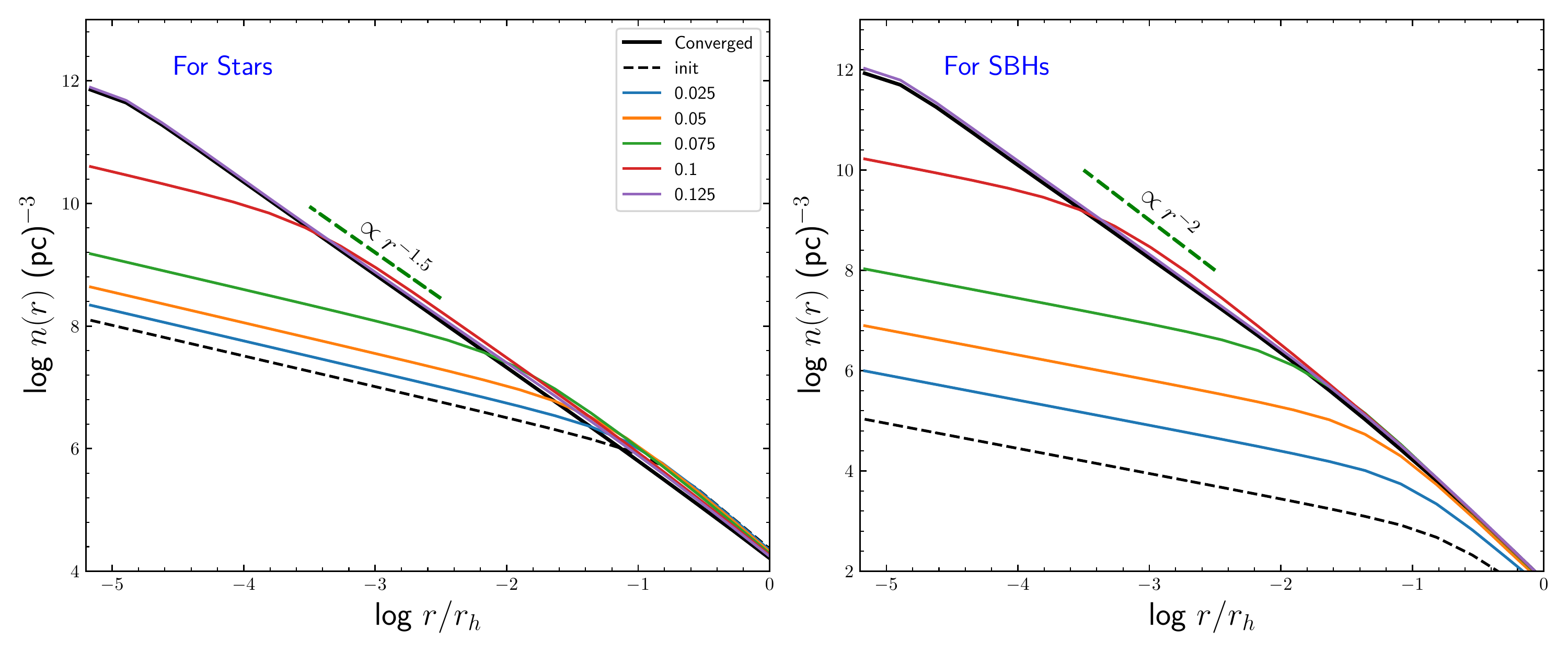}
	\caption{Evolution of the distributions of stars (left panel) and SBHs (right panel) in model M2 as  a
	function of time by \GNC. Loss cone effects are ignored. Different lines are results at different time of simulation, 
	in unit of $T_{\rm rlx}(r_h)$, where $T_{\rm rlx}(r_h)$ is the two-body relaxation time at $r_h$.  
	The mass of the MBH is $\bh=4\times10^6\msun$. 
		}
	\label{fig:relax2}
\end{figure*}

\begin{figure*}
\center
\includegraphics[scale=0.5]{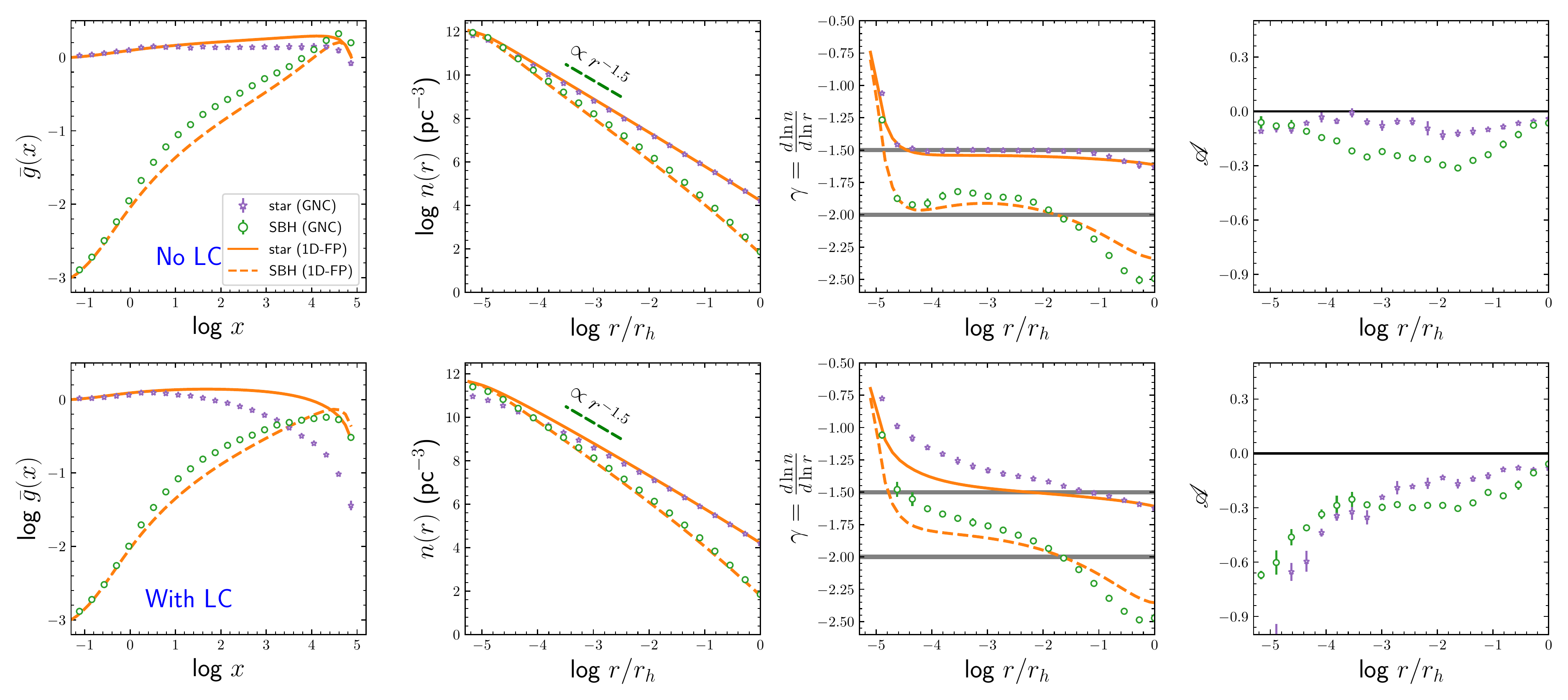}
\caption{Similar to Figure~\ref{fig:fig_gE1}, but for model M2, which consists of stars and SBHs 
(See Table~\ref{tab:model}). }
\label{fig:fig_gE2}
\end{figure*}

\begin{figure*}
	\center
\includegraphics[scale=1.0]{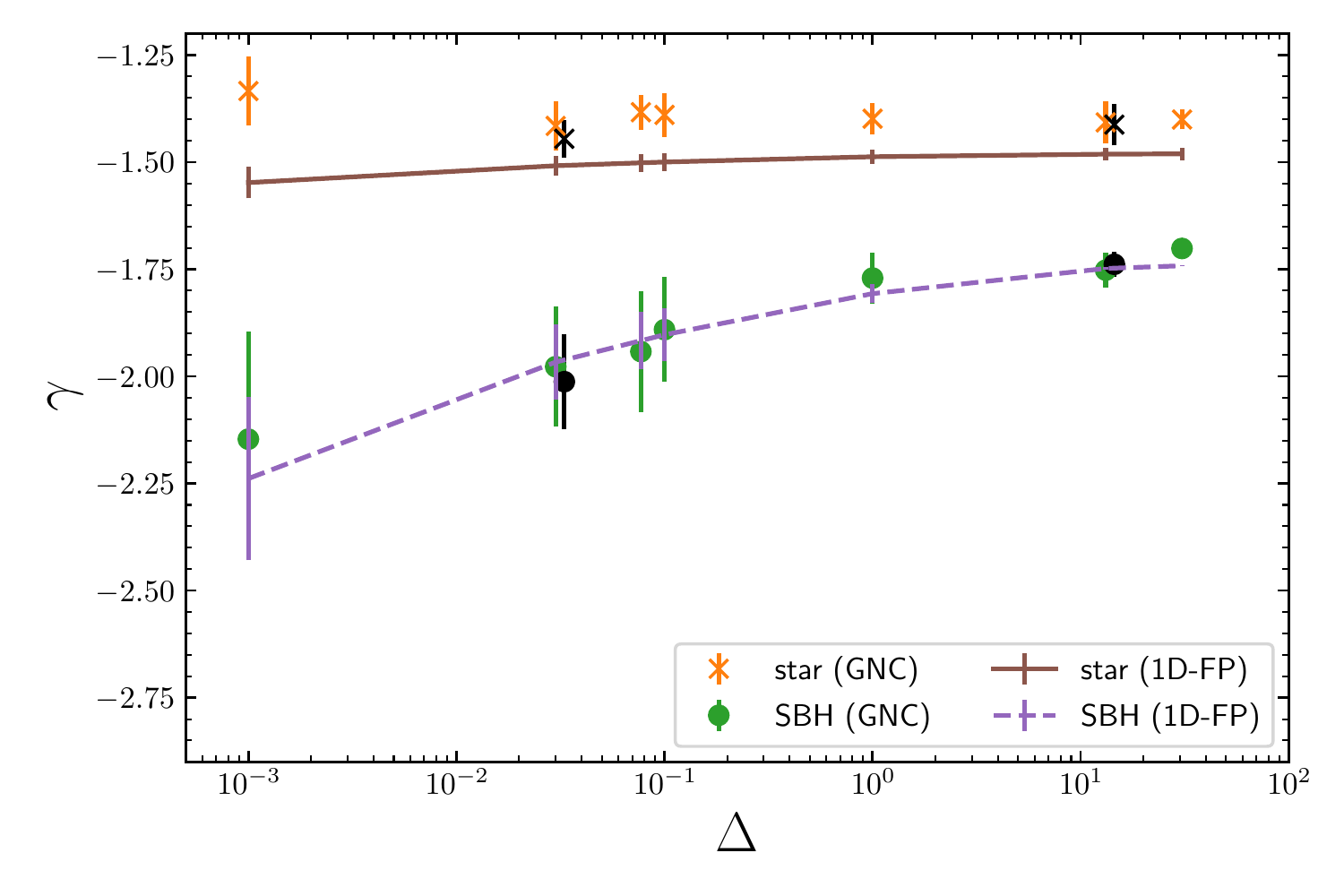}
\caption{The slope index $\gamma=d\ln n(r)/d\ln r$ of density (averaged between $0.001r_h<r<0.1r_h$) as a function 
of the $\Delta$ parameter in model M2 which consists of $1\msun$ stars and $10\msun$ SBHs. 
Those colored lines are results from 1D-FP methods. The  black (or colored) filled circle/cross mark the results from \GNC{} adopting $\bh=10^5\msun$ 
(or $\bh=4\times10^6\msun$).}
\label{fig:fig_slope_delta}
\end{figure*}

\begin{figure*}
\center
\includegraphics[scale=0.5]{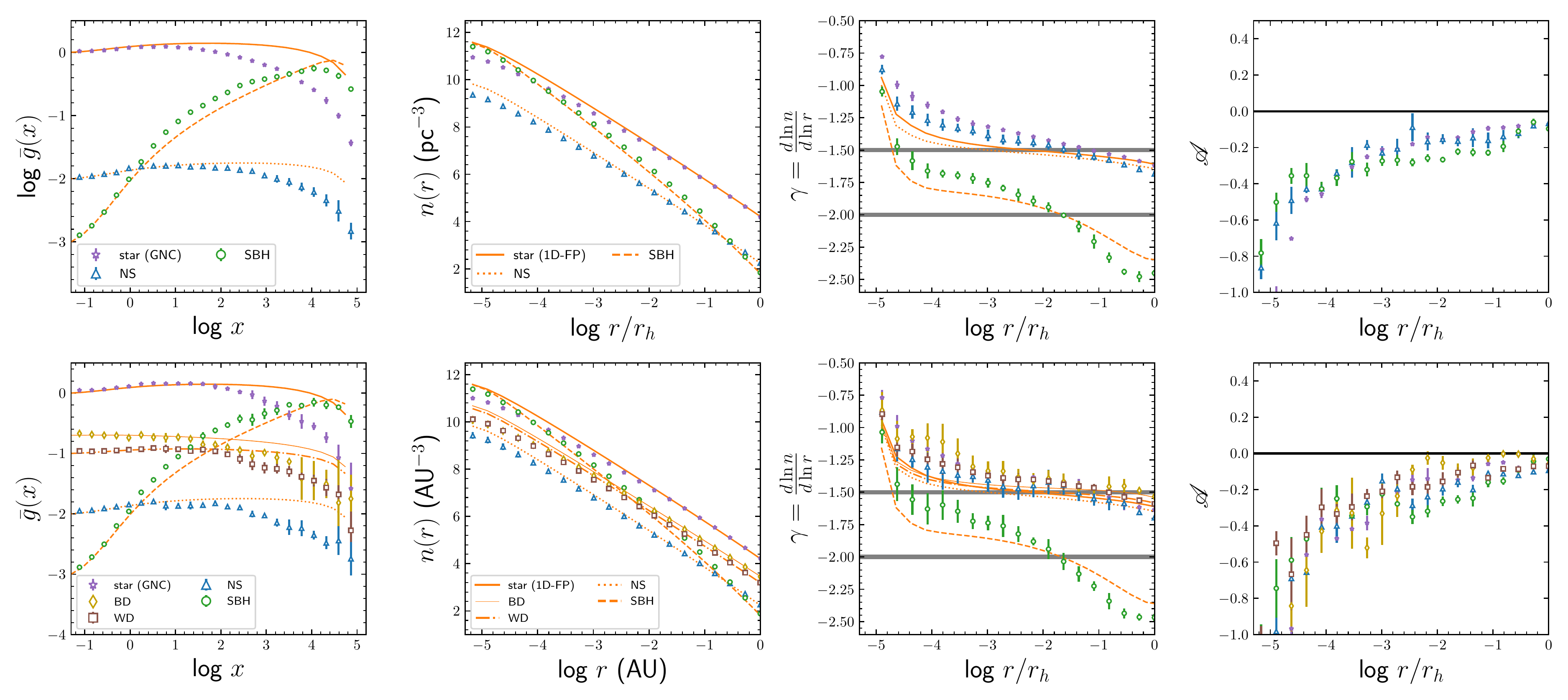}
\caption{Top panels: Similar to Figure~\ref{fig:fig_gE1}, but for the model M3, which consists of stars, neutron stars 
and SBHs; Bottom panels: Similar to the top ones but for the model M5, consisting of 
five components. For details of model M3 and M5 see Table~\ref{tab:model}.
In all panels, the loss cone is considered, symbols are results of \GNC{}, and orange lines 
are results of 1D-FP method. }
\label{fig:fig_gE35}
\end{figure*}

\begin{figure*}
\center
\includegraphics[scale=0.7]{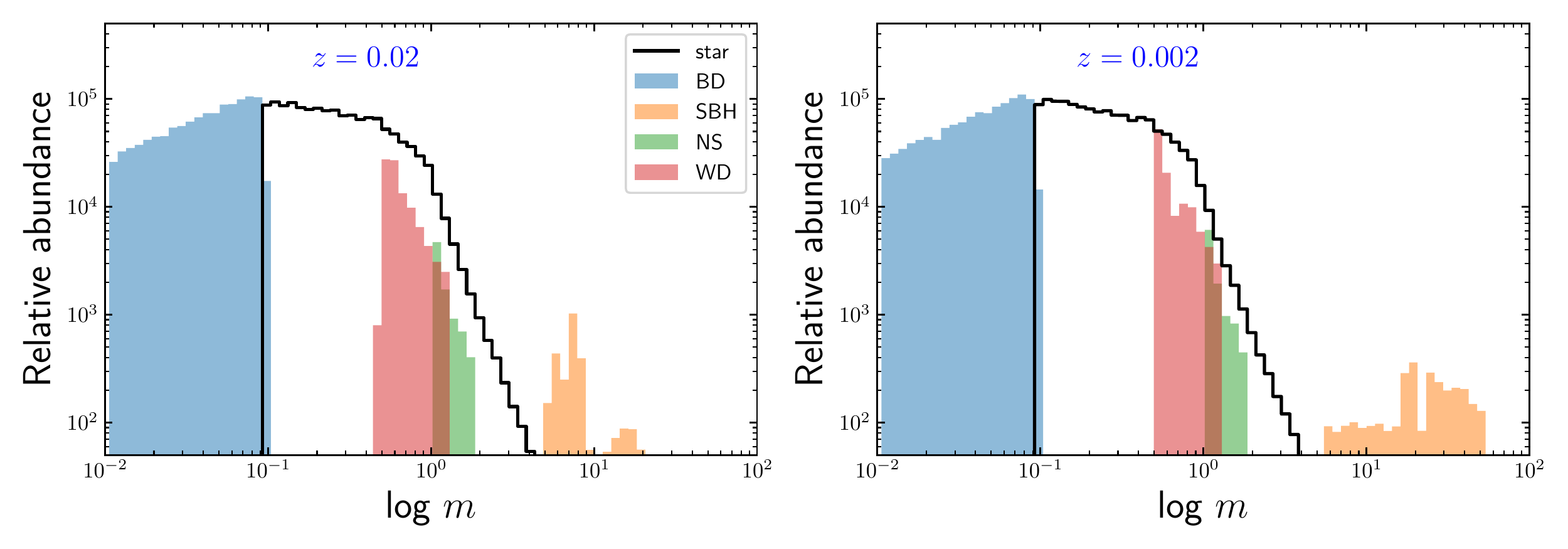}
\caption{Relative abundance of stellar populations from MOBSE after 10 Gyrs of continuous star formation with
Kroupa initial mass function~\citep{2001MNRAS.322..231K}. Left and right panel shows the results 
given metalicity of $z=0.02$ and $z=0.002$, respectively. The asymptotic number ratio of different stellar objects 
at different mass bins are shown in Table~\ref{tab:model_mobse}.}
\label{fig:mf}
\end{figure*}

\begin{figure*}
\center
\includegraphics[scale=1.0]{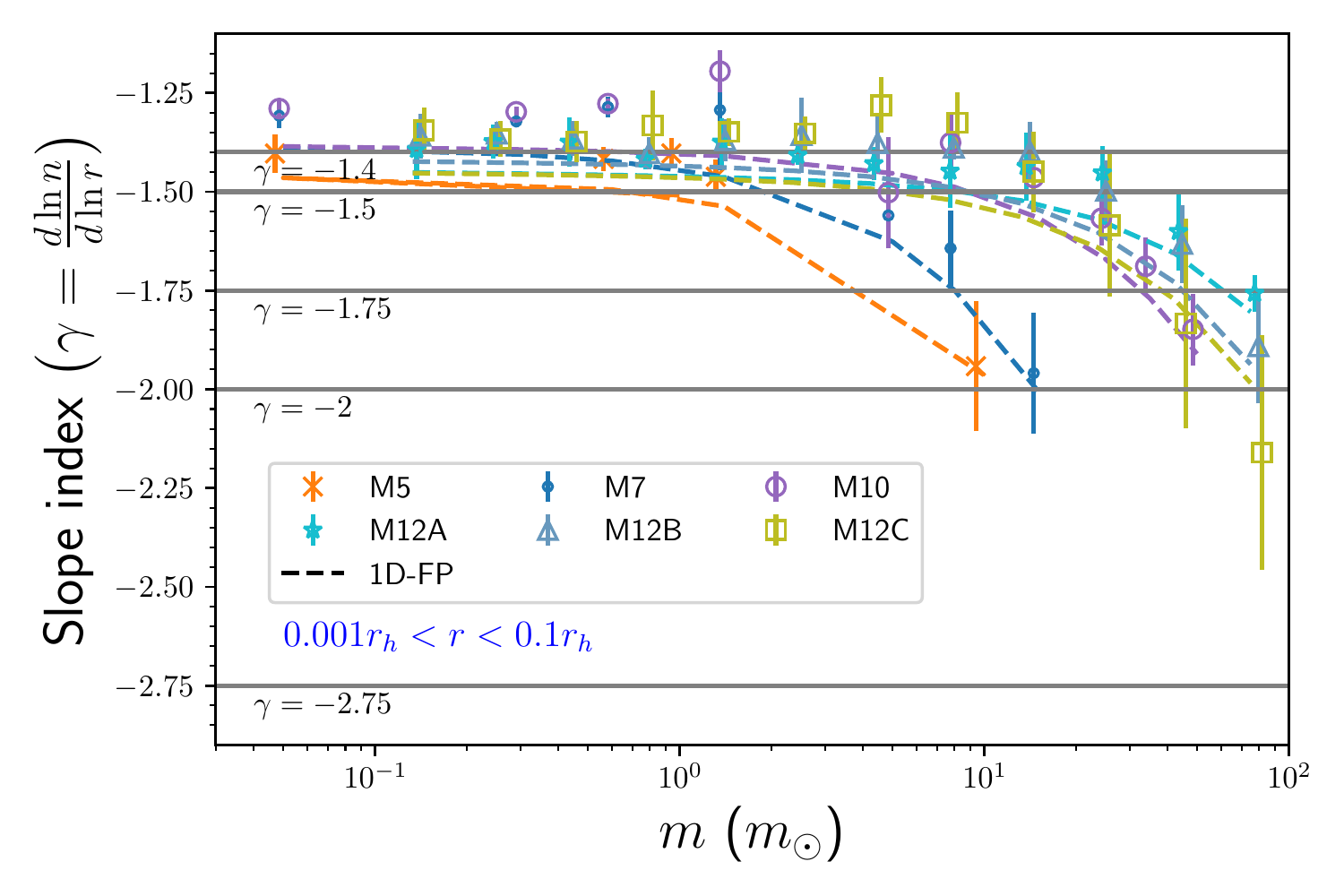}
\includegraphics[scale=1.0]{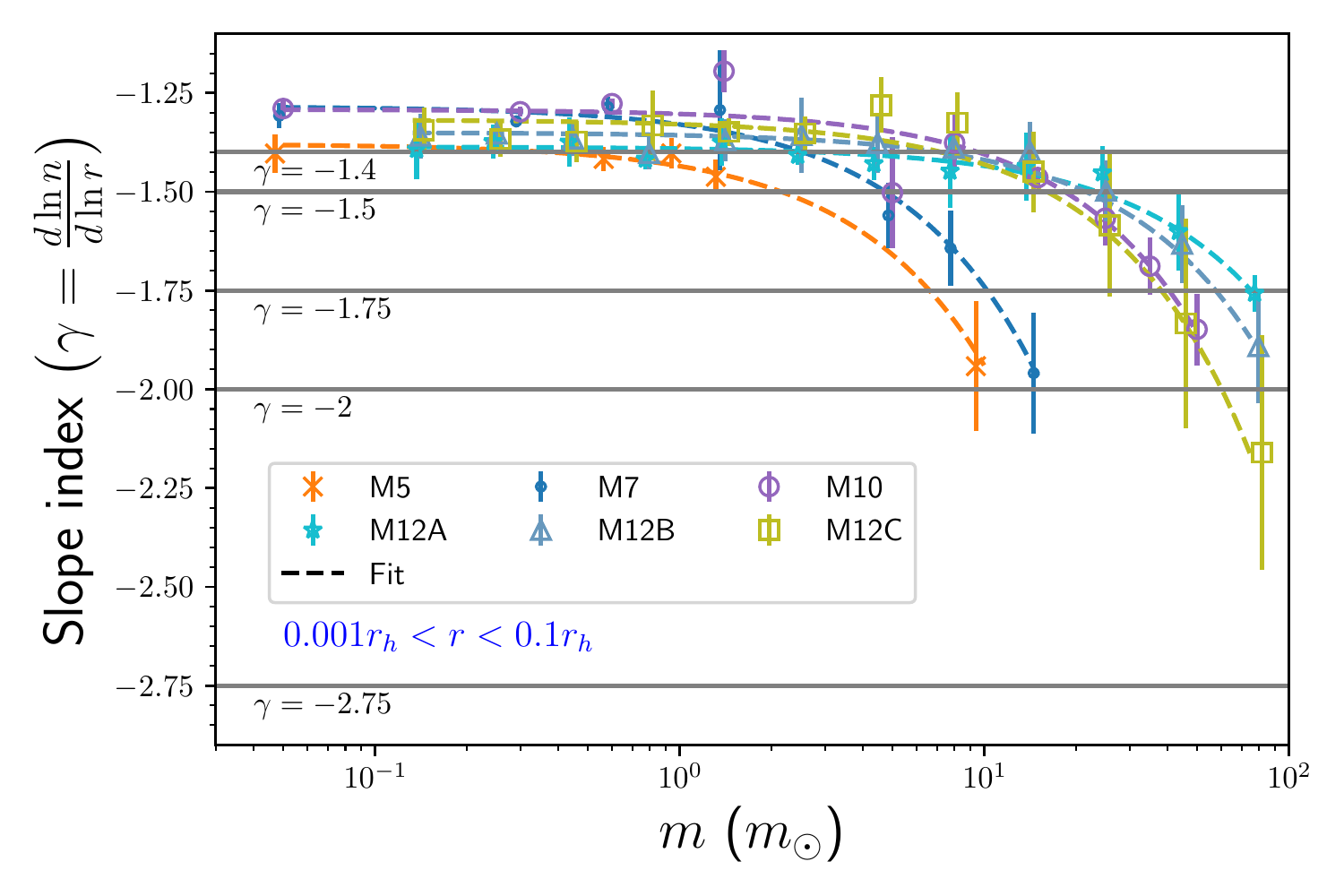}
\caption{Top panel: The symbols represent the slope index $\gamma=d\ln n(r)/d\ln r$ of the density 
(averaged between $0.001r_h<r<0.1r_h$) as a function of the mass of components obtained from \GNC{} 
(including the loss cone). The dashed lines with the same color as the symbols represent the results 
from the 1D-FP method.
Bottom panel: Similar to the top panel, but the dashed lines with the same color as the symbols 
represent the fitting results of $\gamma=-\gamma_0-p_0/(m_{\rm max} m)$, where $\gamma_0$ and $p_0$ 
are free parameters and $m_{\rm max}$ is the maximum mass of the particles. The best-fit values for 
different models can be found in Table~\ref{tab:model_gamma}.
In all panels, the reference lines are as follows: $\gamma=-1.4$, which represents the 
observed density profile of stars in the Galactic center~\citep{2003ApJ...594..812G,2018A&A...609A..27S}; $\gamma=-1.5$, 
which represents the expected index of profiles for light components from 
1D-FP methods in a two-component model~\citep{1977ApJ...216..883B}; $\gamma=-1.75$, 
which represents the slope index for equal-mass star cusps~\citep{BW76}; $\gamma=-2.0$, 
which represents the slope index for stellar-mass black holes 
in a two-component model~\citep{Alexander09,2011CQGra..28i4017A}. The reference line 
$\gamma=-2.75$ represents the slope index expected from the dynamical 
friction limit~\citep{Alexander09}.
Note that the error bar for each symbol represents the variations of $\gamma$ in the range $0.001r_h<r<0.1r_h$.}
\label{fig:fig_slope_idx}
\end{figure*}

When there are multiple mass components in the cluster, the dynamics become more complicated due to mass segregation, which drives a flux of heavy components into the central regions while pushing the light ones out. In this study, we use \GNC{} to investigate the steady-state distribution for systems with multiple (up to 12) mass components.

We start with a two-component model, M2, which consists of stars and SBHs with an asymptotic relative number ratio of $10^{-3}$ at the outer boundary (Table~\ref{tab:model}). The evolution of the density profile for stars and SBHs in M2 from \GNC{} at different times is shown in Figure~\ref{fig:relax2}. Comparing these results to those of model M1 in Figure~\ref{fig:relax}, it is apparent that the relaxation of stars is accelerated due to the presence of SBHs. The relaxation timescale for both stars and SBHs, which both have cavities inside $\sim0.1r_h$, is approximately $0.15T_{\rm rlx}$. These convergence timescales are generally consistent with the regrowth timescale derived by \citet{2011CQGra..28i4017A}.

The steady-state results of M2 from \GNC{} are presented in Figure~\ref{fig:fig_gE2}. We can observe that, if the loss cone is ignored, the slope index of the stars in M2 from \GNC{} is well consistent with $\gamma=-1.5$ across the cluster. When the loss cone effect is included, the density profile of the stars becomes shallower in the inner regions ($\gamma=-1.3\sim-1.4$). The SBHs in M2 concentrate near the center, forming a steeper density profile as expected from mass segregation. The density profile of SBHs varies from $-2.3\sim-1.7$, but can be approximately considered as $\sim-2.0$ in general, for particles between $10^{-3}r_h<r<0.1r_h$.

In model M2, the differences in the density profiles of both stars and SBHs between \GNC{} and the 1D-FP method are quite small, except in the very inner part of the cluster. When the loss cone is considered, the density profiles of the stars from \GNC{} are usually slightly shallower than those in the 1D-FP method. This discrepancy is partly due to the small anisotropy developed in the cluster. The right panels of Figure~\ref{fig:fig_gE2} show the results of the anisotropy for different components obtained from \GNC{}. The maximum tangential anisotropy for all types of objects is approximately $-0.3$ (or $\lesssim-0.6$) when the loss cone is not considered (when the loss cone is considered). For SBHs, the anisotropy is very similar to that of stars.

%We also find consisency on the density profiles of stars and SBHs between those of our method 
%and those of N-body simulations from~\citet{2010ApJ...708L..42P}. The discrepancies are less than $0.3$dex
%for both of the stars and the SBHs. At the outer parts of the cluster the density profile of stars is slightly higher 
%in N-body simulations than those of \GNC{}.

The effects of mass segregation in the case of two component system can be characterized by a $\Delta$ parameter 
introduced by~\citep{Alexander09}
\be
\Delta=\frac{N_H m_H^2}{N_Lm_L^2}\frac{4}{3+m_H/m_L},
\label{eq:delta}
\ee
where $m_H$ ($N_H$) is the mass (asymptotic number density) 
of the heavy component and $m_L$ and $N_L$ is for the lighter component.
We vary the asymptotic number $s_\beta$ at the outer boundary for model M2. The results 
of the slope index of the two components are shown in Figure~\ref{fig:fig_slope_delta}.
The slope index $\gamma$ varies from $-1.75$ to $-2.25$ ($-1.5$ to $-1.6$) 
for the heavy component (light component) when the delta parameter varies 
from $30$ to $10^{-3}$. Thus, the smaller the $\Delta$ parameter, the 
stronger the effect of mass segregation. The trend is generally consistent with 
those of the 1D-FP method calculated in this work, and those from~\citet{Alexander09,2010ApJ...708L..42P}.
However, we notice that 
the density slope of the lighter components from \GNC{} is slightly 
shallower than those of the 1D-FP method. 

We further consider models with more mass components. Figure~\ref{fig:fig_gE35} shows the simulation results of model M3, which includes stars, SBHs, and NSs, and model M5, which additionally includes WDs and BDs. We find that in both of these models, the density profiles of the lighter components (e.g., WDs, BDs, NSs) have a slope index of $-1.3\sim-1.5$, which is very similar to those of the stars. The density profiles of the heavy component, i.e., the SBHs, in these two models are also around $-2.0$, similar to those in model M2.

\subsection{Components with a spectrum of masses }

\begin{table*}%[h]
\caption{Boundary conditions according to MOBSE}
\centering
\begin{tabular}{lcc|cccccc|ccccccccc}
	\hline
	$m_1$  & $m_c$ & $m_2$ &   \multicolumn{6}{c|}{$z=0.02$ (M7)} &  \multicolumn{6}{c}{$z=0.002$ (M10)} \\
	$\msun$ & $\msun$ & ($\msun$) &  $s(m_\alpha)$    & $f_{\rm star}$ & $f_{\rm SBH}$ & $f_{\rm NS}$ & $f_{\rm WD}$  &$f_{\rm BD}$
	& $s(m_\alpha)$    & $f_{\rm star}$ & $f_{\rm SBH}$ & $f_{\rm NS}$ & $f_{\rm WD}$  &$f_{\rm BD}$\\
	\hline
	$0.01$  & $0.05$   & $0.1$ & $0.87$           & $0.0$  & $0.0$ & $0.0$  & $0.0$  & $1.0$ & $0.87$           & $0.0$  & $0.0$ & $0.0$  & $0.0$   & $1.0$\\
	$0.1$  & $0.3$     & $0.5$ & $0.79$           & $1.0$  & $0.0$ & $0.0$  & $0.0$  & $0.0$ & $0.81$           & $1.0$  & $0.0$ & $0.0$  & $0.0$   & $0.0$\\
	$0.5$   & $0.6$    & $1.2$ & $0.27$           & $0.74$ & $0.0$ & $0.0$  & $0.25$ & $0.0$ & $0.27$           & $0.68$ & $0.0$ & $0.0$  & $0.32$  & $0.0$\\
	$1.2$   & $1.4$    & $2.1$ & $0.02$           & $0.73$ & $0.0$ & $0.3$ & $0.0$   & $0.0$ & $0.02$           & $0.6$ & $0.0$ & $0.4$  & $0.0$    & $0.0$\\
%	$2.1$   & $3$      & $4$   & $1\times10^{-3}$ & $1.0$  & $0.0$ & $0.0$ & $0.0$   & $0.0$ & $1\times10^{-3}$ & $1.0$  & $0.0$ & $0.0$  & $0.0$   & $0.0$ \\
	$4$     & $5$      & $6$   & $3\times10^{-4}$ & $0.0$  & $1.0$ & $0.0$ & $0.0$   & $0.0$ & $1\times10^{-4}$ & $0.0$  & $1.0$ & $0.0$  & $0.0$   & $0.0$ \\
	$6$     & $8$      & $10$  & $2\times10^{-3}$ & $0.0$  & $1.0$ & $0.0$ & $0.0$   & $0.0$ & $3\times10^{-4}$ & $0.0$  & $1.0$ & $0.0$  & $0.0$   & $0.0$ \\
	$10$    & $15$     & $20$  & $3\times10^{-4}$ & $0.0$  & $1.0$ & $0.0$  & $0.0$  & $0.0$ & $6\times10^{-4}$ & $0.0$  & $1.0$ & $0.0$  & $0.0$   & $0.0$\\
	$20$    & $25$     & $30$  & $-$  		  	  & $-$    & $-$   & $-$    & $-$    & $-$   & $5\times10^{-4}$ & $0.0$  & $1.0$ & $0.0$  & $0.0$   & $0.0$\\
	$30$    & $35$     & $40$  & $-$ 			  & $-$    & $-$   & $-$    & $-$    & $-$   & $4\times10^{-4}$ & $0.0$  & $1.0$ & $0.0$  & $0.0$   & $0.0$\\
	$40$    & $50$     & $60$  & $-$ 			  & $-$    & $-$   & $-$    & $-$    & $-$   & $4\times10^{-4}$ & $0.0$  & $1.0$ & $0.0$  & $0.0$   & $0.0$\\	
	\hline
	%& 0.99 & 45$\arcdeg$ & 180$\arcdeg$  \\ \hline
	%
	\end{tabular}
\tablecomments{The asymptotic number ratio at the outer boundary ($s_\beta$$(m_\alpha)=s(m_\alpha)f_\beta$) of model M7 and M10, for 
different kinds of stellar objects and mass bins. These numbers are simulation results from MOBSE and corresponds to the relative abundance 
shown in Figure~\ref{fig:mf}.
}
\label{tab:model_mobse}
\end{table*}

\begin{table*}%[h]
\caption{Fitting of the slope index of density profiles for multiple mass component models}
\centering
\begin{tabular}{l|ccc|ccc|ccccc}
	\hline
	\multirow{2}{*}{Model}  & $m_{\rm min}$ &  $m_{\rm max}$ & \multirow{2}{*}{$\Delta$} &
	\multicolumn{3}{c|}{1D-FP} & \multicolumn{3}{c}{\GNC} \\
	& ($\msun$) & ($\msun$) &  &$p_0$ & $\gamma_0$ &$100\eta$ &  $p_0$ & $\gamma_0$ & $100\eta$\\
	\hline
	M5    & $0.05$ & $10$& $0.029$   &$0.51$ & $1.47$ & $5.1$  & $0.56$ & $1.37$ & $5.7$\\
	M7    & $0.05$ & $15$& $0.16$    &$0.62$ & $1.40$ & $4.1$  & $0.68$ & $1.28$ & $4.5$\\
	M10   & $0.05$ & $50$& $0.57$    &$0.53$ & $1.39$ & $1.1$  & $0.57$ & $1.29$ & $1.1$\\
	M12A  & $0.13$ & $75$& $91$      &$0.36$ & $1.46$ & $0.48$ & $0.35$ & $1.38$ & $0.47$\\	
	M12B  & $0.13$ & $75$& $1.5$     &$0.55$ & $1.39$ & $0.74$ & $0.51$ & $1.35$ & $0.69$\\	
	M12C   &$0.13$ & $75$ & $0.028$  &$0.54$ & $1.47$ & $0.72$  & $0.85$ & $1.37$ & $1.1$\\	
	\hline
	%& 0.99 & 45$\arcdeg$ & 180$\arcdeg$  \\ \hline
	%
	\end{tabular}
\tablecomments{The fitting results of the slope index as a function of mass from \GNC{} or 1D-FP method to 
the function $\gamma=-\gamma_0-\eta m=-\gamma_0-p_0/m_{\rm max}m$. 
$\Delta$ parameter is estimated by Equation~\ref{eq:delta2}.\\
}
\label{tab:model_gamma}
\end{table*}	

We also tested the profiles for systems that consist of components with a broad spectrum of masses in \GNC{}. First, we tested a simple system with a power-law mass function for stars, which was designed for testing purposes. In the $\alpha$-th mass bin $m_\alpha$, the asymptotic number ratio is given by
\be
s_\star(m_\alpha) =C_\alpha m_\alpha^{\gamma_{\star}}
\ee
where
\be
C_\alpha=m_\alpha \log(10) d (\lg m) \frac{m_{\rm max}^{1+\gamma_\star}-m_{\rm min}^{1+\gamma_\star}}{1+\gamma_\star}
\ee

We use $12$ mass bins range from $0.1-100\msun$, each of which centered at $0.133$, $0.237$, $0.422$, 
$0.750$, $1.33$, $2.37$, $4.22$, $7.50$, $13.3$, $23.7$, $42.2$, and $75.0$ $(\msun)$. 
Each mass bin is composed by stars only, for simplicity. 
We test three models with $\gamma_\star=-1.3$, $-2.3$ and $-3.3$, which is named model M12A, M12B and M12C, respectively. 
%Note that in the model M12C, the asymptotic number ratios $s_\star$ vary from $0.734$ to $3.5\times10^{-7}$ when %masses vary from $0.1\msun$
%to $\sim75\msun$, thus, the most massive bin has less than one (real) particles. However, 
%the simulated number of particles in the most massive bin 
%can be increased to over hundreds of thousands by setting the weighting of particles to be $w_n=0.01$
%described in Section~\ref{subsec:weighting}. Similarly, by 
%using the method, we can have enough number of particles for statistics in all mass bins.

In the case of systems with a broad spectrum of masses, it has already been found that the steady-state number density of 
particles in each mass bin follows its own power-law profiles $n (r)\propto r^{\gamma_m}$~\citep{1977ApJ...216..883B, 
2009ApJ...698L..64K,2006ApJ...649...91F}.
1D-FP methods similar to those in Appendix~\ref{subsec:analy_method} found that~\citep{1977ApJ...216..883B, Oleary09}
\be
\gamma_m=-\gamma_0-\eta m=-\gamma_0-\frac{p_0}{m_{\rm max}}m
\label{eq:gammam}
\ee
where $\gamma_0=3/2$, $p_0$ is a parameter mainly determined by mass segregation and $m_{\rm max}$ is the maximum mass among all 
bins. The $\Delta$ parameter defined by Equation~\ref{eq:delta} now becomes~\citep{Alexander09}
\be
\Delta=\frac{4\langle N m^2\rangle_H}{3\langle N m^2\rangle_L+\langle m_H\rangle\langle Nm\rangle_L},
\label{eq:delta2}
\ee
where the components are separated into high and low mass groups if specifying a given boundary of mass. 
Usually  the boundary of mass is $5\msun$~\citep{Alexander09}. 
1D-FP method suggests that, if $\Delta \ga 0.1$, $p_0=0.25\sim0.3$~\citep{1977ApJ...216..883B,Alexander09}. 
If $\Delta\lesssim0.1$, $p_0$ will increase with the $\Delta$ parameter~\citep{Alexander09}.

For all models we adopt $5\msun$ as the boundary separating the low and high mass groups.
The $\Delta$ parameters for different models are shown in Table~\ref{tab:model_gamma}.
We fit simultanously Equation~\ref{eq:gammam} to the slope index results of \GNC{}, 
and the best fitting value of $\gamma_0$ and $p_0$ parameter for different models are shown in Table~\ref{tab:model_gamma}. 

Figure~\ref{fig:fig_slope_idx} show the $\gamma_m$ as a function of $m$ in different models from \GNC{}.
In most models the slope index of light components ($\lesssim 5\msun$) varies very slowly around $-1.3\sim-1.4$.
We notice that the slope index of stars  is consistent with those of the observered stars 
at the Galactic center~\citep[$\sim-1.4$,][]{2003ApJ...594..812G,2018A&A...609A..27S}. 

For heavy components ($\ga 5\msun$) the density profile is steeper for more massive particles. 
In most models (expect M12C) the slope index of the most massive bin is between $-1.75$ and $-2.0$. 
For weak mass segregation ($\Delta>0.5$, e.g., in model M12A), 
$p_0\sim0.35$, which is slightly larger than $\sim0.3$ expected from the 1D-FP method
~\citet{1977ApJ...216..883B}. When the mass segregation is strong ($\Delta\lesssim 0.5$), 
$p_0$ can be $0.5\sim0.9$ and apparently increased with the parameter $\Delta$ and is related to the 
the maximum mass of particles. 

Model M12C have the strongest mass segregation effects among all models, 
where the most heavy component ($75\msun$) has a density slope of $-2.2\sim-2.3$, which is close to the 
expected limit of dynamical friction ($=-11/4$)~\citep{Alexander09}. 

Table~\ref{tab:model_gamma} and Figure~\ref{fig:fig_slope_idx} suggest that, even in the case of a broad spectrum of masses, 
both results from \GNC{} and 1D-FP are in general consistent with 
each other, although we find that the 1D-FP usually over predict the density of particles 
in the inner parts of the cluster, compared with those from \GNC{}.

In reality, however, the mass function of particles, including stars and compact objects, is a slightly complex function and 
does not strictly follow a continuous power-law profile. For a more realistic treatment of the boundary condition, here we 
use MOBSE~\citep{2018MNRAS.474.2959G,2018MNRAS.480.2011G}\footnote{Although MOBSE can handle binary evolutions, in this work it is only used to evolve single stars.} 
to generate stellar populations after a continuous star formation last for $10$Gyr.
We use Kroupa initial mass function~\citep{2001MNRAS.322..231K} for stars between $0.01$ and $150\msun$.
The results are shown in Figure~\ref{fig:mf} and the asymptotic number ratio 
of each mass bin is listed in Table~\ref{tab:model_mobse}. We test two different models, 
named M7 and M10, of which adopt metalicity $Z=0.002$ and $Z=0.02$, respectively. 

The slope index as a function of masses can be also found in Figure~\ref{fig:fig_slope_idx}. 
The $\Delta$ parameter of M7 and M10 is given by $0.16$ and $0.57$ (see Table~\ref{tab:model_gamma}), respectively,
suggesting that the mass segregation in M7 is relatively stronger than those in M10. 
The power-law index for SBHs is around $-2.0$ for M7 (at $\sim 10\msun$) and 
 around $\sim -1.8$ for M10 (at $\sim 50\msun$). 
These results suggest that in real cluster system the $\Delta$ parameter is $\sim0.1-0.6$, and that 
the mass segregation is modest such that 
the slope index of the most heavy single component is about $-1.8\sim-2.0$.

%{\color{red} test $\sigma_M^2=G\bh/(1+\alpha)r$???}

\subsection{Consumption rates of particles in the loss cone}
\label{subsec:flux}
\begin{figure*}
\center
\includegraphics[scale=0.5]{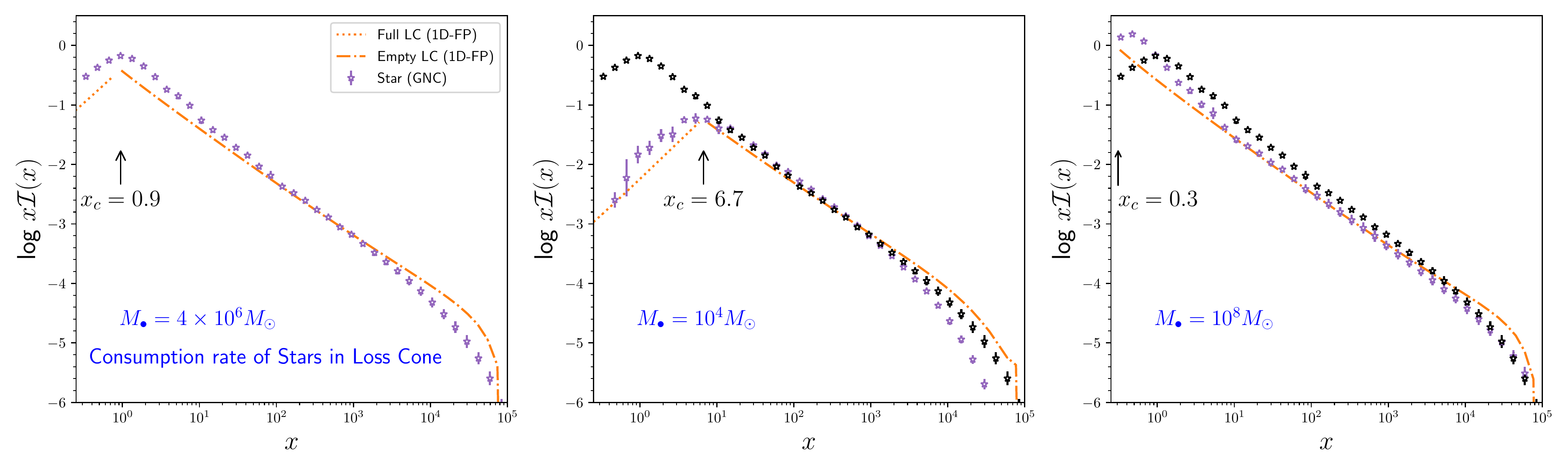}
\caption{The dimensionless flux $x\mathcal{I}(x)$ (per log $x$, given by Equation~\ref{eq:Ix}) 
of stars to the loss cone for  model (M1) which consists of equal mass of stars. The mass of MBH is $\bh=4\times10^6\msun$ (left panel), 
$10^4\msun$ (middle panel) and $10^8\msun$ (right panel). 
%The physical flux per unit energy $E$ is given by $F_{{\rm lc},\alpha}(E)=F_0\mathcal{I}(x)$ 
%and $F_0$ is given by Equation~\ref{eq:F0}.
Symbols are results of \GNC. The dotted (or dashed) orange lines show the analytical results from 1D-FP method 
in the case of full (or empty) loss cone, according to Equation~\ref{eq:flc1} (or Equation~\ref{eq:flc2}). 
The arrow marks the transition position ($x_c$) according to Equation~\ref{eq:analy_qxc}. The black stars in 
the middle and right panel are the results of \GNC{} given $\bh=4\times10^6\msun$ (the stars in the 
left panel), which is used for comparison. }
\label{fig:fig_flux}
\end{figure*}
	
\begin{figure}
\center
\includegraphics[scale=0.5]{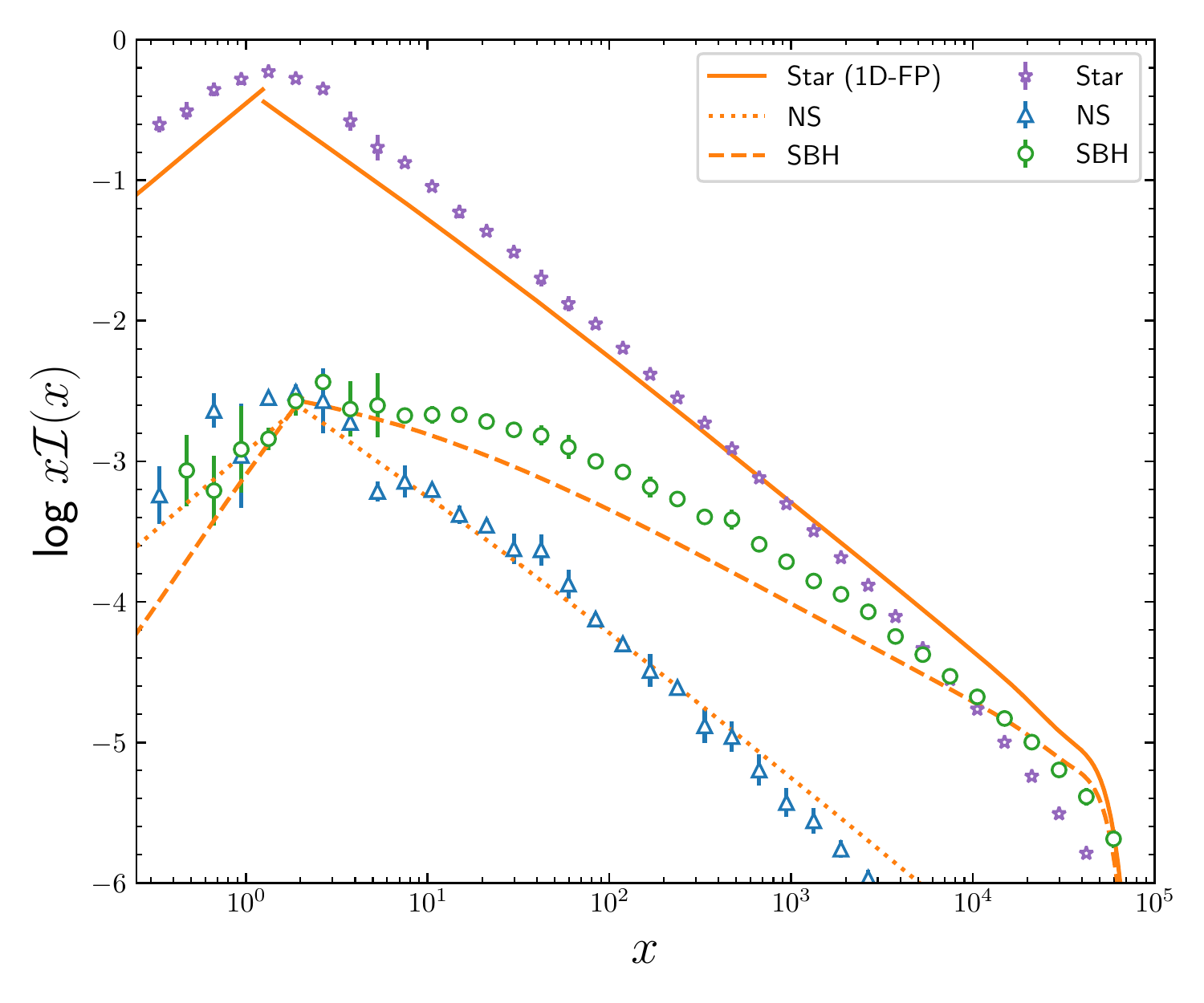}
\caption{Similar to Figure~\ref{fig:fig_flux}, but the dimensionless flux of 
stars, NSs and the SBHs in model M3. }
\label{fig:fig_flux2}
\end{figure}
Stellar objects will be destroyed if they are wandering too close to the MBH. In our simulations this happens when
particles move inside the loss cone, and that the particles must have passed through the pericenter of the orbit. 
In steady state there is a constant flux of particles that vanish 
inside the loss cone. In this section we examine the flux of different kinds of particles vanished in the loss cone by both \GNC{} and the 1D-FP method. 

Left panel of Figure~\ref{fig:fig_flux} show the results of the dimensionless flux $x\mathcal{I}(x)$ 
(per log $x$) for  model M1 with $\bh=4\times10^6\msun$. The flux at $x\ll x_c$
are expected to be different with those at $x\gg x_c$, where $x_c\simeq0.9$ ($\simeq 0.56r_h$) (Equation~\ref{eq:analy_qxc}), of which the former follows Equation~\ref{eq:flc1}, 
and the later Equation~\ref{eq:flc2}.
We can see that there is a general consistency between the results of \GNC{} and those from the 1D-FP method. 
The difference are within a factor of $\sim2$, and slightly larger at the inner and outer parts of the cluster. 

We notice that 1D-FP method assumes that $\bar g(x)$ follows a power-law profile, and that the 
angular momentum distribution of particles is isotropic, which are both not entirely true in \GNC. 
The discrepancy at inner regions is possibly because 1D-FP method over predict the number of particles 
at that region (See the left panel in Figure~\ref{fig:fig_gE1}). 
The loss cone at the outer part of the cluster ($x\sim 1$) is not either  
full ($q\gg1$) nor empty ($q\ll1$) 
(for $\bh=4\times10^6\msun$), thus the predicted flux rates from 1D-FP method 
is less accurate.

We also investigate the consistencies of these two methods for MBHs with different masses.
For a given type of stellar object, the size of the loss cone, and also the consumption rate 
in the loss cone vary with the mass of MBH. The transition dimensionless energy 
 $x_c$ decrease with the mass of MBHs. As shown in Figure~\ref{fig:fig_flux}, we find general consistency between \GNC{} and the 1D-FP method
for MBHs with masses range from $10^4\msun$ to $10^8\msun$. We notice that the dimensionless flux of empty 
loss cone appears in a  self-similar way above the transition energy, 
although the mass of MBH is different. 

We also examine the consumption rates for models consist with multiple mass components and different stellar types. 
Figure~\ref{fig:fig_flux2} show the results of the flux of stars, NSs and SBHs in model M3. 
We find that 1D-FP method also have general consistencies with \GNC{} for all mass components. 
The transition energy and the amount of consumption rates of stars are weakly affected by 
the presence of other components. As the loss cones of NSs and SBHs are much smaller than those of stars, 
the position of transition is deeper, which is located at $x_c\sim 1.7$ ($a_c\sim 0.3r_h$). It seems that the 
1D-FP method slightly lower estimates the consumption rates of SBHs at the empty loss cone region by about a 
factor of $<2\sim 3$, compared to those from \GNC. 

The above estimations are for stellar populations bound to the cluster. The total consumption rate  can then be obtained by 
summing both the bound and unbound populations 
(Equation~\ref{eq:R_tdu}). Note that the consumption rates of bound 
population will be larger than those of unbound one when $x_c\ga 1.6$. For stars, that happens when the mass of MBH is $\le 10^6\msun$. 

If assuming a Milky-Way MBH, the consumption rate from \GNC{} is 
 $\sim 1.3\times10^{-4}$ yr$^{-1}$ for bound stars (or $\sim 2.5\times10^{-4}$ yr$^{-1}$ for both bound and unbound). 
The rate estimated by 1D-FP method is  $\sim 8.2\times10^{-5}$ yr$^{-1}$ for bound stars
(or $\sim 1.9\times10^{-4}$ yr$^{-1}$ for both bound and unbound).
The difference of these two methods is about $20-30\%$. 
These values are in general consistent with those of other authors~\citep[e.g.,][]{1999MNRAS.309..447M,
2004ApJ...600..149W}. 

\section{Discussion and conclusions}
In this work, we have developed a Monte Carlo method (\GNC) to study the dynamical evolution of multiple mass components in a star cluster containing a massive black hole at its center. The basic version of the method presented here incorporates two-body relaxation and the effects of the loss cone. For the first time, we calculate the two-body relaxation of multiple mass components based on a two-dimensional (energy and angular momentum) Fokker-Planck (FP) Monte Carlo method, allowing us to obtain steady-state solutions for multiple mass components under given boundary conditions.

We find that \GNC{} yields results that are consistent with those obtained from one-dimensional FP (1D-FP) methods. This consistency holds for various models consisting of equal or multiple mass components and for black holes spanning different mass scales. We also investigate the anisotropy of the cluster and observe the development of tangential anisotropy in the inner regions, while the outer parts can remain isotropic. This anisotropy is the primary cause of the discrepancies between our method and 1D-FP methods, as the latter assume isotropy of the cluster.

Nevertheless, it is important to emphasize that \GNC{} is a Monte Carlo method that offers flexibility and advantages over both 1D-FP methods and N-body simulations. Some of these advantages include:

The ability to incorporate complex dynamics in a straightforward manner, such as resonant relaxations, stellar collisions, tidal dissipations, gravitational wave dissipations of orbits, complex dynamics of binaries, and more, while simultaneously considering the two-dimensional evolution of particles in energy and angular momentum.

Also, the ability to increase the number of particles in the simulation arbitrarily using weighting methods. This is particularly useful for rare objects such as stellar black holes or stellar binary black holes, allowing us to obtain better statistical results for these particles.

The capability to achieve well-converged density profiles and accurate particle evolution down to distance scales of $10^{-5}r_h$ from the black hole using particle splitting and weighting methods. This level of accuracy is challenging to achieve with N-body simulations, which typically approach $0.01-0.001r_h$. Accurate particle evolution in these regions is crucial for studying phenomena such as tidal disruption of stars and extreme-mass-ratio inspirals (EMRIs).

However, it is important to acknowledge some limitations of our current version of the method that need to be addressed in future studies. The most significant limitation is the absence of the stellar object potentials in estimating particle energy and angular momentum. This omission introduces inaccuracies in the density profile and flux into the loss cone from our method, particularly in regions beyond $r_h$. As a result, the results in the outer parts of the cluster are only approximate. Additionally, due to this limitation, we can currently only consider the steady state of the cluster by implementing outer boundary conditions. In the future, when we include the effects of stellar potentials, we will be able to obtain the time-dependent evolution of the cluster and the mass of the black hole.

Nevertheless, our method remains accurate below regions of $0.1r_h$, where the potential of the black hole dominates over that of the stars. Therefore, we can still apply our method to study various interesting events in galactic nuclei driven by dynamical processes occurring in the inner regions of the cluster. Examples of such events include partial or full tidal disruption of stars of different types and masses, as well as the properties of different types of gravitational wave sources, both first-generation and multiple-generation sources. These applications are particularly relevant for future multi-messenger observations in galactic nuclei.
%\begin{acknowledgments}
\section{Acknowledgments}
\noindent
This work was supported in part byNational Natural Science Foundation of 
China under grant No. 12273006, 11603083, U1731104, 
the Natural Science Foundation of Guangdong Province under grant No. 2021A1515012373. 
This work was also supported in part by the Key Project of the National Natural Science Foundation 
of China under grant No. 11733010,12133004. 
The simulations in this work are performed partly in the TianHe II National
Supercomputer Center in Guangzhou. We acknowledge the funds from the ``European Union NextGenerationEU/PRTR'', Programa de Planes Complementarios I+D+I (ref. ASFAE/2022/014).
%\end{acknowledgments}
%-----------------------------------------------------------------------

\appendix
\section{Monte-Carlo simulation of a particle}
\label{apx:MC_method}
\begin{figure*}
	\center
	\includegraphics[scale=0.75]{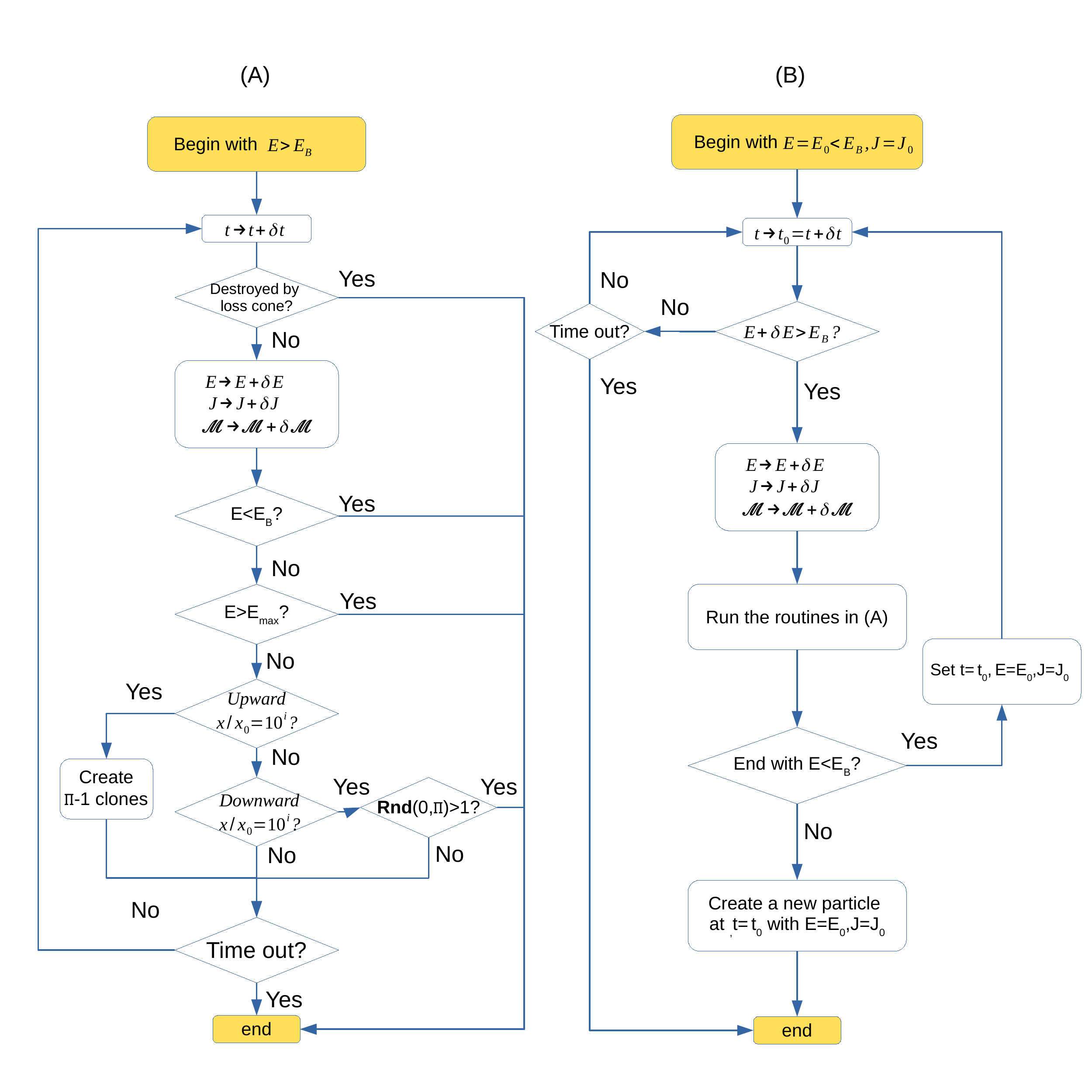}
	\caption{Flow charts depicting the subroutines of particles inside the boundary (left panel) 
	and those outside the boundary (right panel). $\Pi$ is the clone factor for particles crossing 
	the dimensionless energy position $x/x_0=10^i$, $i=1,\cdots, 4$, where $x=E/\sigma_h^2$, and $x_0$ is an arbitrary energy above which the clone scheme is implimented. 
	 }
	\label{fig:flowchart}
\end{figure*}

In Monte-Carlo simulations, particles outside the boundary $E<E_B$
($E_B=x_B\sigma_h^2$), which do not move into the cluster in the next timestep
($E<E_B\rightarrow E<E_B$), should be considered as fixed and not move in $E-J$
spaces. This is because the outer boundary, by definition, represents the
Dirichlet boundary condition of Equation~\ref{eq:EJequation}. Thus, it is
equivalent to a scheme where every particle that has changed its energy and
angular momentum ($E$ and $J$) values is immediately replaced by a new particle
with the same $E$ and $J$ values, ensuring that the distribution of particles
on the boundary remains unchanged. However, in the case where a particle
successfully moves into the cluster by crossing the boundary ($E<E_B\rightarrow
E>E_B$) at a specific time $t=t_0$, to compensate for the void created by the
injection of the particle, a new particle with the exact same previous values
of $E$ and $J$ should be created at that time. By repeating this process for
all particles, there will be a constant flow of particles injected into the
cluster from the outer boundary.

On the other hand, if a particle is located inside the boundary $E_B$ and
attempts to move outside of it ($E>E_B\rightarrow E<E_B$) in the next step, its
simulation should be stopped. In other words, the particle is eliminated at
that boundary. The injection of particles from the outer boundary and the
elimination of particles that flow out of it will gradually balance each other,
ultimately creating a smooth distribution of particles near the outer boundary.

According to the described schemes, the Monte-Carlo simulations of particles in \GNC{} proceed as follows, one-by-one in sequence:

If the particle is outside the boundary of the cluster, with $E=E_{0}<E_B$ and $J=J_{0}$, then it follows this process:

\begin{enumerate}
    \item Obtain the changes $\delta E$ and $\delta J$, as well as $\delta t$ using the functions provided in Appendix~\ref{apx:delta_functions}. Update $t$ by setting $t\rightarrow t+\delta t$.
    \item If $E+\delta E>E_B$, proceed to Step 3. Otherwise, save $t_0=t$ and continue to Step 4.
    \item If the simulation time has reached the limit, terminate the simulation. Otherwise, return to Step 1.
    \item Update $E$ by setting $E\rightarrow E+\delta E$, $J\rightarrow J+\delta J$, and $\mathscr{M}\rightarrow \mathscr{M}+\delta \mathscr{M}$.
    \item Perform the Monte-Carlo steps for particles inside the cluster, as described below.
    \item If the particle ends up with $E<E_B$, set $t=t_0$, $E=E_0$, and $J=J_0$, then return to Step 1. Otherwise, proceed to the next step.
    \item Create a new particle at time $t=t_0$, $E=E_0$, and $J=J_0$.
    \item Terminate the simulation for the current particle.
\end{enumerate}

If a particle is within the cluster, the simulation runs in the following way:

\begin{enumerate}
    \item Calculate the time step, $\delta E$, $\delta J$, and the change in mean anomaly $\delta \mathscr{M}$ using the functions provided in Appendix~\ref{apx:delta_functions}. Update $t$ by setting $t\rightarrow t+\delta t$.
    \item If $J<J_{\rm lc}$, where $J_{\rm lc}$ is the loss cone angular momentum, test whether the mean anomaly of the orbit has passed through the pericenter. If $\mathscr{M}<\pi$ and $\mathscr{M}+\delta \mathscr{M}>\pi$, or if $\delta t>P$, the particle is considered destroyed by the loss cone, and the simulation for that particle stops. If not, proceed to the next step.
    \item Save $E_i=E$, $J_i=J$, then update $E$ by setting $E\rightarrow E+\delta E$, $J\rightarrow J+\delta J$, and $\mathscr{M}\rightarrow \mathscr{M}+\delta \mathscr{M}$. Save $E_f=E$, $J_f=J$.
    \item If $E<E_B=x_B\sigma_h^2$, the simulation for the particle stops because it moves outside the outer boundary.
    \item If $E>E_{\rm max}=x_{\rm max}\sigma_h^2$, the simulation for the particle stops because it moves outside the inner boundary.
    \item If $E_{i}/(x_0\sigma^2_h)<10^i$ and $E_{f}/(x_0\sigma^2_h)>10^i$, create $\Pi-1$ clone particles.
    \item If $E_{i}/(x_0\sigma^2_h)>10^i$ and $E_{f}/(x_0\sigma^2_h)<10^i$, and if ${\rm RND}(0,\Pi)>1$, the simulation for the particle stops. Otherwise, proceed to the next step.
    \item If the simulation time has reached the limit, the simulation stops. Otherwise, return to Step 1.
\end{enumerate}

The details of the particle simulation in these two cases are also illustrated in Figure~\ref{fig:flowchart}.

In our simulation, the dimensionless angular momentum of particles can range
from $j_{\rm min}$ to $1$. If $j$ moves inside $j_{\rm min}$ (or outside $1$),
we apply $j\rightarrow 2j_{\rm min}-j$ (or $j\rightarrow 2-j$), as $D_{EE}$
diverges near $j\rightarrow 0$. The value of $j_{\rm min}$ should be much
smaller than the minimum size of the loss cone for all types of particles in
the simulation. In most cases, it is sufficient to set $j_{\rm min}$ to a value
between $10^{-4}$ and $10^{-3}$.

\section{Time steps and evolution in two-body relaxation}
\label{apx:delta_functions}

The time step $\delta t$ of simulation for a particle required by the two-body relaxation is given by~\citep{SM78}
\be\ba
\delta t\le \delta t_E&={\rm min}\left[\frac{(0.15E)^2}{D^{EE}(E,J)}, \left|\frac{0.15E}{D^{E}(E,J)}\right|\right]\\
\delta t\le \delta t_J&={\rm min}\left[\frac{(0.1J_c)^2}{D^{JJ}(E,J)}, \frac{[0.4(1.0075J_c-J)]^2}{D^{JJ}(E,J)}\right]
\ea\ee

If the loss cone effect is additionally considered, and the size of the loss cone is $J_{\rm LC}$, then the time steps need to be adjusted according to the following requirement:

\be\ba
\delta t\le \delta t_{\rm LC}&=\frac{{\rm max}\left[0.1J_{\rm LC}, 0.25(J-J_{\rm LC})\right]^2}{D^{JJ}(E,J)},\\
\ea\ee

If the particle is within the loss cone and has not been previously destroyed by it in a previous step, we additionally require that the next time step cannot be larger than the period of the orbit:

\be
\delta t\le P(E).
\ee
where $P(E)$ is the orbital period of the particle.

Given the time step, we evolve the $E$, $J$ and $\mathscr{M}$ according to two-body relaxation
\be\ba
\delta E&=D^E\delta t+y_1\sqrt{D^{EE}\delta t}\\
\delta J&=D^J\delta t+y_2\sqrt{D^{JJ}\delta t}\\
\delta \mathscr{M}&=\frac{2\pi }{P(E)}\delta t
\ea\ee
where $y_1$ and $y_2$ are two unit normal random numbers with correlation $\rho=\frac{D^{EJ}}{\sqrt{D^{EE}D^{JJ}}}$.

Note that in this work, we only consider the effects of two-body relaxation on
the orbits to test the basic version of \GNC. It is straightforward to add the
effects of resonant relaxation, gravitational wave dissipation, and other
factors in each time step. These aspects will be discussed in our future works.

\section{Monte-Carlo simulations and steady state solutions}
\label{apx:MC_steady}
We take the following steps to get the steady-state distributions of particles:
\begin{enumerate}
\item
Initially,  we generate a number
\be
N_\beta(m_\alpha)=\frac{\bh }{m_\star w_\alpha}s_\beta(m_\alpha)
\ee

\noindent 
of particles of type $\beta$ with mass $m_\alpha$,
where $w_\alpha$ is the weighting factor of the particles in mass bin $\alpha$ (See Section~\ref{subsec:weighting}), 
and $s_\beta(m_\alpha)$ is the number ratio of the particle beyond the 
outer boundary (See Section~\ref{subsec:boundary_condition}).

The initial energy distribution $E$ of all particles follows a guessed
distribution $N(E)\propto E^{\alpha_{\rm E,ini}-2}$ (or $N(a_2)\propto
a_2^{\alpha_{\rm E, ini}}$) between $0.03<x<100x_B$, where $\alpha_{\rm E,ini}$
ranges from $0$ to $0.25$. The distribution of particles in the boundary
regions ($0.03<x<x_B=0.05$) is determined by the value of $\alpha_{\rm E,ini}$.
However, as long as the thickness of the boundary is sufficiently small, the
model results are insensitive to the specific value of $\alpha_{\rm E,ini}$.
Changing it to other values usually does not lead to noticeable differences in
the model results.
The angular momentum $j$ of each particle at the boundary follows the
distribution given by $G(j)$ in Equation~\ref{eq:gj}. Initially, a random mean
anomaly between $0$ and $2\pi$ is assigned to the outer orbit $\mathscr{M}$ for
each particle.

\item We calculate the number distribution in the mass bin $\alpha$, $N_\alpha(E,J)$, and the dimensionless functions $g_\alpha(x,j)$ and $\bar g_\alpha(x)$ according to Equation~\ref{eq:galpha} and Equation~\ref{eq:bar_galpha}, respectively.
We then calculate the weighting constant $w_n$ by normalizing $\bar g_\alpha(x_B)$ (usually selected from the first mass bin, $\alpha=1$) in Equation~\ref{eq:galpha_norm}. This normalization is typically necessary before the third or fourth iterations. After that, the weighting $w_n$ remains constant as the distributions of particles near $x_B$ converge.

\item We calculate the tables of diffusion coefficients shown in Equation~\ref{eq:dedj1}
and~\ref{eq:dedj2} with size of $96\times96$ (or $120\times120$) in the space
of $\log x-\log j$. 

\item Run the Monte-Carlo subroutines for each particle as described in
Appendix~\ref{apx:MC_method}. Run the simulation for a duration that is about a
few fractions of the two-body relaxation time at $r_h$, typically between
$0.005$ and $0.01$ times $T_{\rm rlx}(r_h)$.  During the simulation, obtain the
diffusion coefficients at any given values of $x$ and $j$ through interpolation
or finding the nearest value from the table prepared in the previous step. Keep
track of particles that exit normally (exit due to reaching the end of the
simulation time).  To reduce Monte-Carlo errors, refresh the angular momentum
of each particle outside the boundary before the next iteration by replacing
$j$ with a new value that follows the distribution $G(j)$.  Repeat the process
from step 2 until the profiles of $\bar g_\alpha(x)$ in all mass bins have
converged.

\end{enumerate}

Similar to~\citet{1978ApJ...226.1087C}, to increase the computational efficiency we can generate the following table of 
an auxiliary function $C(s,e)$ that can be used for all simulations:
\be
C^{lmn}(s,e)=2^{1-l}\int^{{\rm min}\left[1,\left(\frac{2}{s}-1\right)\frac{1}{e}\right]}_{-1}
\frac{(1+ye)^{l+n}}{(1-ye)^{n+m/2}\sqrt{1-y^2}}(2-s-sye)^{m/2}dy
\ee

Then the $\Gamma$ functions in Equation~\ref{eq:dedj1} and~\ref{eq:dedj2} can be fastly integrated by 
\be
\Gamma^{lmn}_\beta(E, e)=\frac{\kappa}{\pi}\int^{\frac{2}{1-e}}_{1}\bar f_\beta(sE) C^{lmn}(s,e)ds
\ee
where $\kappa=16 \pi^2G \ln\Lambda$.

Our computational cost is approximately proportional to the number of particles
and slightly increases with the number of mass bins. The most computationally
expensive case in our simulation is model M12A, which takes around 30 hours to
simulate 1 million particles with 12 mass bins (as discussed in
Section~\ref{subsec:multiple_masses}) for a duration of approximately one
relaxation timescale ($T_{\rm rlx}(r_h)$). This computation is performed on an
Intel Xeon CPU with a clock speed of 2.20 GHz, utilizing 24 threads.  For the
other models in this work, the simulation times typically range from one to
several hours. We generally run the simulations for a duration of at least 1.5
times $T_{\rm rlx}(r_h),$ although the profiles usually converge after
approximately 0.3-0.4 times $T_{\rm rlx}(r_h)$.

\section{Review of the one-dimensional FP method in calculating the relaxation- and loss-cone effects}
\label{subsec:analy_method}

In one-dimensional FP methods, the evolution of particles primarily focuses on
the evolution of energy and assumes that the angular momentum distribution of
any particle in the $\alpha$-th mass bin is always isotropic, i.e.,
$f_\alpha(E,J)=\bar f_\alpha(E)$~\citep{BW76}. Consequently, the $j$-averaged
diffusion coefficients of the $\alpha$-th bin reduce
to~\citep{1977ApJ...216..883B}:

\be\ba
\bar D^{E}_\alpha &=\frac{\int f_\alpha(E,J) D^{E}_{\alpha}(E,J) dJ}{\int f_\alpha(E,J) dJ}=\frac{\int 2J D^{E}_\alpha(E,J) dJ}{J^2_{c}}\\
=&8 A \sum_\beta \left[m_\alpha m_\beta
\int_E^\infty\frac{\bar f_\beta(E')}{(2E')^{5/2}}dE'-m_\beta^2\int^E_{-\infty}\frac{\bar f_\beta (E')}{(2E)^{5/2}}dE'\right]\\
\bar D^{EE}_\alpha&=\frac{\int f_\alpha(E,J) D^{EE}_{\alpha}(E,J) dJ}{\int f_\alpha(E,J) dJ}=\frac{\int 2J D^{EE}_\alpha(E,J) dJ}{J^2_{c}}\\
=&\frac{16}{3} A \sum_\beta\left[m_\beta^2
\int_E^\infty\frac{\bar f_\beta(E')}{(2E')^{3/2}}dE'+m_\beta^2\int^E_{-\infty}\frac{\bar f_\beta(E')}{(2E)^{3/2}}dE'\right]\\
\label{eq:javgdedee}
\ea\ee
where $A=2\sqrt{2}\pi G\ln \Lambda E^{5/2}=2\sqrt{2}\pi G\ln \Lambda x^{5/2}\sigma_h^5$, $\Lambda=\bh/m_\alpha$. 

Under the assumption of isotropic angular momentum distribution, we 
substitute Equation~\ref{eq:javgdedee} in Equation~\ref{eq:EJequation}, 
and then substitute $\bar f(E)$ by the dimensionless distribution $\bar g(x)$, 
finally the equation reduces to~\citep{1977ApJ...216..883B} 
%\be\ba
%\frac{\pl f}{\pl t}=-\frac{\pl R(E)}{\pl E}
%\ea\ee
%where
%\be\ba
%R(E)=\frac{32}{3}2^{1/2}\pi^5\ln\Lambda G^5 \bh^3\sum_\beta m_\alpha m_\beta \int^\infty_{-\infty}
%\frac{1}{{\rm max}(E,E')^{3/2}}\left(\frac{\pl f_\beta (E')}{\pl E'}-\frac{m_\beta}{m_\alpha}
%\frac{\pl f_\beta (E)}{\pl E}f(E')\right)dE'
%\ea\ee
%Or in a more simple form given by~\citet{2006ApJ...645L.133H} 
\be
\frac{\pl \bar g_\alpha(x)}{\pl \tau}=-x^{5/2}\frac{\pl Q_\alpha(x)}{\pl x}-\mathcal{F}_{{\rm lc},\alpha}(x)
~\label{eq:ana1d}
\ee
where 
\be\ba
Q_\alpha(x)&=\sum_\beta m_\alpha m_\beta \int_{-\infty}^{x_{\rm max}}dy \left[\bar g_\alpha (x)\frac{\pl \bar g_\beta(y)}{\pl y}\right.\\
&\left.-\frac{m_\beta}{m_\alpha}\bar g_\beta(y)\frac{\pl \bar g_\alpha(x)}{\pl x}\right]{\rm max}(x,y)^{-3/2}
\ea\ee
and
\be
\tau =\frac{t}{\tau_0},~~\tau_0=\frac{3}{32\pi^2}\frac{(2\pi\sigma_h^2)^{3/2}}{(G m_\star)^{2}\ln \Lambda n_h}
\ee
where $\tau_0$ is about half of the relaxation time at $r_h$.

The loss cone effect can be incorporated into Equation~\ref{eq:ana1d} by adding
an additional term $\mathcal{F}_{{\rm lc},\alpha}(x)$ that describes the
consumption rates of objects in the loss cone. The value of $\mathcal{F}_{{\rm
lc},\alpha}(x)$ depends on a dimensionless parameter $q=D_{JJ,0}P(E)/J_{\rm
lc}^2$, where $D_{JJ,0}=D_{JJ} (J\rightarrow0)\simeq 2JD_J$. This parameter
distinguishes between the empty ($q\ll1$) and full ($q\gg1$) loss cone regions.
The term $\mathcal{F}_{{\rm lc},\alpha}(x)$ is related to the physical flux
$F_{{\rm lc},\alpha}(E)$ (per unit time $t$ and unit $E$) by the following
equation:

\be
F_{{\rm lc},\alpha}(E)dE=
A(2\pi\sigma_h^2)^{-3/2}n_h \tau_0^{-1} \mathcal{F}_{{\rm lc},\alpha}(x)\sigma_h^2 dx
=F_0  \mathcal{I}(x)dx, 
\label{eq:flc_physics}
\ee
where we have defined 
\be
\mathcal{I}(x)= \mathcal{F}_{{\rm lc},\alpha}(x) x^{-5/2},
\label{eq:Ix}
\ee
which is a dimensionless flux and $F_0$ is given by
\be
F_0= \frac{4\sqrt{2}\pi^2}{3} r_h^3 n_h^2 \ln \Lambda \frac{(G m_\star) ^2}{\sigma_h^3}.
\label{eq:F0}
\ee
According to the physical flux given by~\citet{1977ApJ...211..244L}, we have
\be
\ba
F_{{\rm lc},\alpha}(E)&= q \frac{4\pi^2J_{\rm lc}^2\bar f_\alpha(E)}{\ln J_c/J_{\rm lc}},~q\ll1\\
&=1.442\pi^2J_{\rm lc}^2 \bar f_\alpha(E),~q\gg1.
\label{eq:flc_phy}
\ea\ee
Substituting the above equation into Equation~\ref{eq:flc_physics}, we obtain that, when $q\ll1$,
\be
\mathcal{F}_{{\rm lc},\alpha}(x)\simeq\frac{\bar g_\alpha(x)}{\ln J_c(x)/J_{\rm lc}}
\sum_\beta \left(\frac{m_\beta}{m_\star}\right)^2 \bar g_\beta (x), ~~q\ll1
\label{eq:flc1}
\ee
which is exactly the Equation 6 of~\citet{2006ApJ...645L.133H}. 

If $q\gg1$, similarly, we have
\be\ba
\mathcal{F}_{{\rm lc},\alpha}(x)&=\frac{0.54}{\pi^{3/2}}x^{5/2}\frac{\bar g_\alpha (x)}{n_hr_h^3\ln \Lambda}\frac{r_t}{r_h}\frac{\bh^2}{m_\star^2},~~q\gg1\\
&\simeq \frac{1}{10.3\eta} x^{5/2} \bar g_\alpha(x),~q\gg1
\label{eq:flc2}
\ea\ee
where 
\be
\eta=n_hr_h^3\frac{r_h}{r_t}\left(\frac{m_\star}{\bh}\right)^2\ln\Lambda
\ee

Note that when deriving Equation~\ref{eq:flc1} and~\ref{eq:flc2}, it is
necessary to make the assumption that $\bar g_\beta (x)$ (for all mass bins
$\beta$) scales with some power of
$x$~\citep{1977ApJ...211..244L,1977ApJ...216..883B}, even though the actual
form of $\bar g_\beta (x)$ may appear arbitrary in these equations. This
assumption introduces some uncertainties in the flux rates, typically resulting
in a factor of approximately 2 variation in most cases.

We can then obtain the expression of the parameter $q$:
\be
q(x)=\frac{D_{JJ}P(E)}{J_{\rm lc}^2}
\simeq 3.7\times \eta x^{-5/2} \sum_\beta\left(\frac{m_\beta}{m_\star}\right)^2\bar g_\beta(x),
\label{eq:analy_q}
\ee

In full loss cone regions ($q\gg1$), $N_\alpha(E,J)\propto J$~\citep{1977ApJ...211..244L} 
and thus the number distribution of angular momentum given $x$ is given by
\be
N(x,j)\propto j
\label{eq:gj_fulliso}
\ee 
For empty loss cone regions ($q\ll1$), for a particle of type $\beta$, 
the distribution is given by~\citep{1977ApJ...211..244L, 1978ApJ...226.1087C, Baror16} 
\be
N_\beta(x,j)\propto \frac{\ln(j/j_{\rm lc,\beta})}{\ln (1/j_{\rm lc,\beta})-0.5}
\label{eq:gj_empiso}
\ee

We define $j_{\rm lc,\beta}=J_{\rm lc,\beta}/J_c=\sqrt{2 r_{t,\beta}/a_2}$ 
as the dimensionless size of loss cone for the objects of type $\beta$, so that it can be described by 
\be\ba
j^2_{\rm lc,\beta}&=\frac{J^2_{\rm lc,\beta}}{J_c^2}\\
&=1-\left(1-\frac{r_{\rm td}}{a_2}\right)^2.
\ea\ee

For stars, $r_{\rm td}=R_\star\left(\frac{3\bh}{m_\star}\right)^{1/3}$ where 
$R_\star=0.00465$AU is the solar radius; For compact objects, 
such as WD, NS and SBH, $r_{\rm td}=8r_g$ (for a Newtonian loss-cone), where
$r_g=\frac{G\bh}{c^2}$. For other type of objects such as binaries, their loss 
cone can be obtained in a similar fashion.

Equation~\ref{eq:flc1} and~\ref{eq:flc2} intersects 
roughly at $x_c$ where $10.3\simeq  - 3.7 \ln j_{\rm lc}(x_c) q^{-1}(x_c)$, thus 
\be
q(x_c)\simeq -0.36 \ln j_{\rm lc}(x_c).
\label{eq:analy_qxc}
\ee
For various type of objects (normal stars or compact stars), and a range of MBH masses ($10^4\msun<\bh<10^8\msun$), 
usually $q(x_c)$ is about orders of $\sim 1$, and $x_c\sim 1-10$. 

The total rate of particles into the loss cone is then 
\be
R_{\rm td}=F_0\int_{x_{\rm min}}^{x_B}\mathcal{I}(x)x d\ln x
\label{eq:R_td}
\ee

We notice that the above consumption (loss) rates are for particles bound to the cluster. According to the~\citet{SM78}, 
the consumption rate of unbound population is given by 
\be
R_{\rm td}^u\simeq \frac{0.2R_{\rm td}\ln (x_{\rm max}/(4x_c))}{x_c^{5/4}}.
\label{eq:R_tdu}
\ee
Then the total consumption rate can be obtained by summing up these two 
different terms
\be
R_{\rm td}^t\simeq R_{\rm td}\left[1+\frac{0.2\ln (x_{\rm max}/(4x_c))}{x_c^{5/4}}\right].
\ee
Note that given $x_{\rm max}=10^5$, the consumption rates of bound 
population will dominate over those of unbound ones when $x_c\ga 1.6$.

\end{document}